\RequirePackage[2020-02-02]{latexrelease}
\documentclass[twocolumn,prl]{revtex4}

\usepackage{ulem}

\usepackage{graphicx}
\usepackage{amsmath}
\usepackage{amssymb}
\usepackage{color}

\newcommand{\RefTitle}[1]{``#1,''}
\newcommand{\SM}[1]{{\color{black} #1}}
\newcommand{\SMexp}{\SM{Sec.~I of the Supplemental Material (SM)}}
\newcommand{\SMeq}{\SM{Sec.~II of the SM}}
\newcommand{\SMddiA}{\SM{Eqs.~(S15) and (S16) in the SM}}
\newcommand{\SMldk}{\SM{Eq.~(S18) in the SM}}

\newcommand{\SMexpANDrb}{\SM{Secs.~I and IV of the SM}} 
\newcommand{\SMslp}{\SM{Sec.~V of the SM}}
\newcommand{\SMddi}{\SM{Sec.~VI of the SM}}

\newcommand{\SMphase}{\SM{Sec.~VIII of the SM}}
\newcommand{\SMbec}{\SM{Sec.~IX of the SM}}

\begin{document}

\title{Experimental Demonstration of Stationary Dark-State Polaritons Dressed by Dipole-Dipole Interaction}

\author{
Bongjune Kim,$^{1,*}$
Ko-Tang Chen,$^{1}$
Kuei-You Chen,$^{1}$
Yu-Shan Chiu,$^{1}$
Chia-Yu Hsu,$^{1}$
Yi-Hsin Chen,$^{2,3}$
Ite A. Yu$^{1,3,}$}\email{upfe11@gmail.com; yu@phys.nthu.edu.tw}

\affiliation{
$^{1}$Department of Physics, National Tsing Hua University, Hsinchu 30013, Taiwan \\
$^{2}$Department of Physics, National Sun Yat-sen University, Kaohsiung 80424, Taiwan \\
$^{3}$Center for Quantum Science and Technology, National Tsing Hua University, Hsinchu 30013, Taiwan
}

\begin{abstract}
Dark-state polaritons (DSPs) based on the effect of electromagnetically induced transparency are bosonic quasiparticles, representing the superpositions of photons and atomic ground-state coherences. It has been proposed that stationary DSPs are governed by the equation of motion closely similar to the Schr\"{o}dinger equation and can be employed to achieve Bose-Einstein condensation (BEC) with transition temperature orders of magnitude higher than that of the atomic BEC. The stationary-DSP BEC is a three-dimensional system and has a far longer lifetime than the exciton-polariton BEC. In this work, we experimentally demonstrated the stationary DSP dressed by the Rydberg-state dipole-dipole interaction (DDI). The DDI-induced phase shift of the stationary DSP was systematically studied. Notably, the experimental data are consistent with the theoretical predictions. The phase shift can be viewed as a consequence of elastic collisions. In terms of thermalization to achieve BEC, the $\mu$m$^2$-size interaction cross-section of the DDI can produce a sufficient elastic collision rate for the stationary DSPs. This work makes a substantial advancement toward the realization of the stationary-DSP BEC.
\end{abstract}

\maketitle

\newcommand{\FigOne}{
\begin{figure}[t]
	\includegraphics[width=0.85\columnwidth]{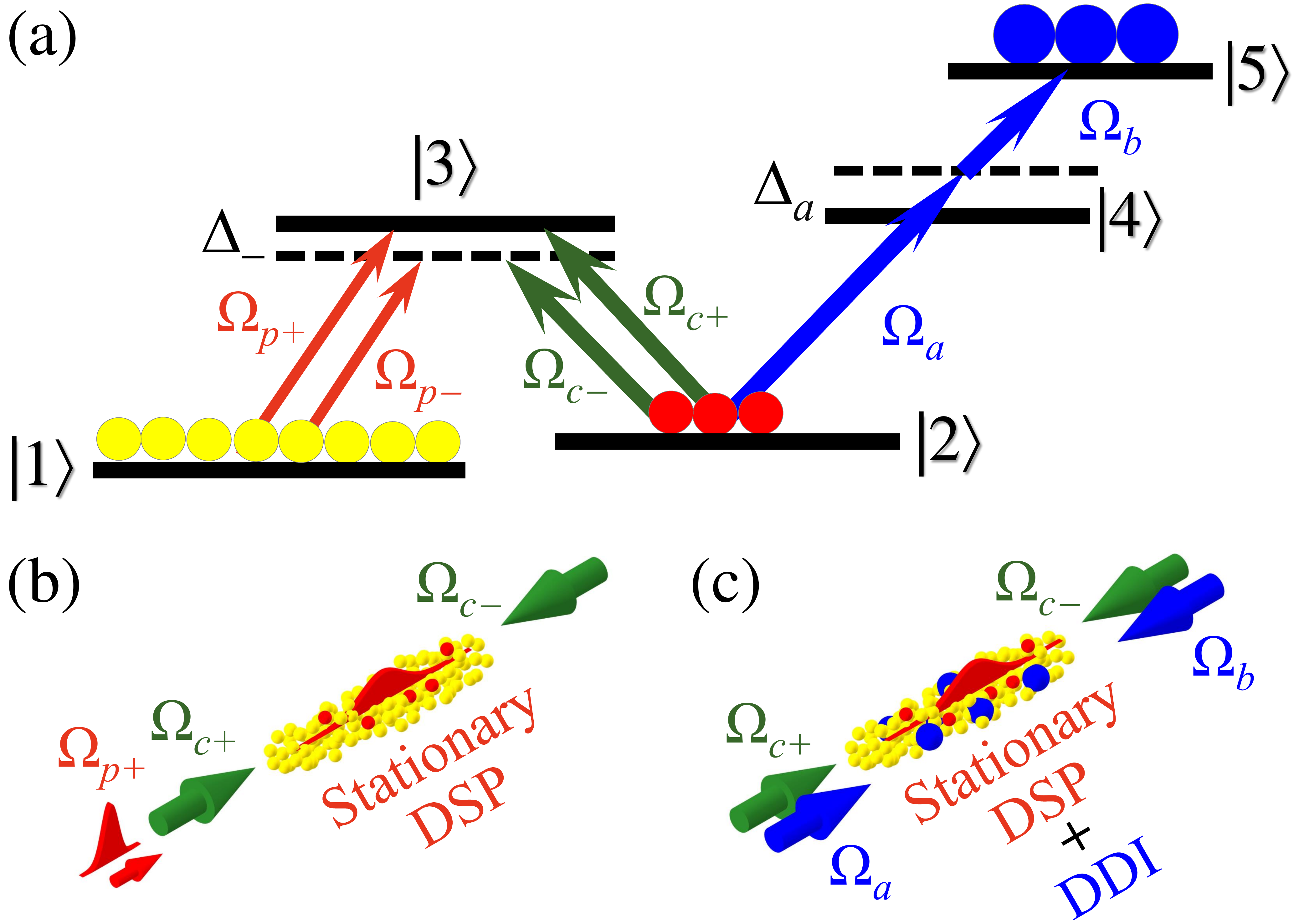}
	\caption{(a) Relevant energy levels and laser excitations. States $|1\rangle$, $|2\rangle$, and $|3\rangle$ form the $\Lambda$-type EIT system for the creation of stationary DSPs, where $\Omega_{p\pm}$ and $\Omega_{c\pm}$ denote the Rabi frequencies of the forward/backward probe and coupling fields, respectively, and $\Delta_- =$ $-1$$\Gamma$. Population oscillates between $|2\rangle$ and a Rydberg state $|5\rangle$, driven by the two-photon transition (TPT) of the fields $\Omega_a$ and $\Omega_b$ with $\Delta_a = +5$$\Gamma$, where $|4\rangle$ is an intermediate state of the TPT. The spontaneous decay rates of $|3\rangle$ and $|4\rangle$ are about the same and denoted as $\Gamma$ (= 2$\pi$$\times$6~MHz). In this work, the pair of $\Omega_{c+}$ and  $\Omega_{p+}$, that of $\Omega_{c-}$ and  $\Omega_{p-}$, and that of $\Omega_a$ and  $\Omega_b$ always maintained the two-photon resonances. (b,\! c) Propagation directions of the laser fields in the measurements of stationary DSPs without and with the TPT or equivalently the DDI. 
	}
	\label{fig:transition}
	\end{figure}
}
\newcommand{\FigTwo}{
\begin{figure*}[t]
	\includegraphics[width=\textwidth]{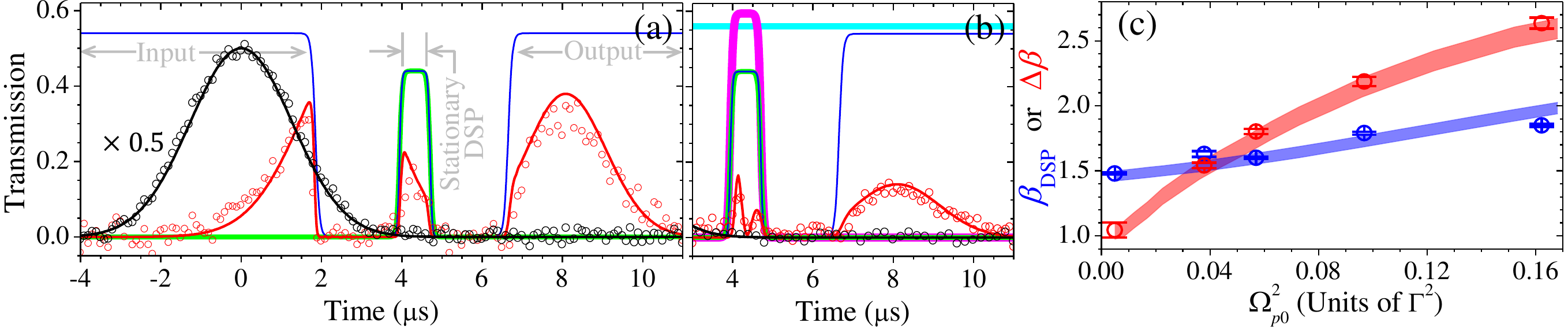}
	\caption{
(a,\! b) 
Representative data demonstrate the consistency with predictions. The analysis of data and the illustration of parameters can be found in text and Sec.~III of the SM. Black and red circles are the data of the input and output probe pulses in the forward direction. Black line is the Gaussian best fit of the input probe pulse, and red lines are the predictions of the output probe pulses. Blue, green, magenta, and cyan lines represent the timing sequences of the forward and backward coupling fields $\Omega_{c+}$ and $\Omega_{c-}$ and the TPT fields $\Omega_a$ and $\Omega_b$. We do not plot the data before $t <$ 3 $\mu$s in (b), which are the same as those in (a). The peak Rabi frequency of the input probe pulse, $\Omega_{p0}$, is 0.07$\Gamma$. The experimental parameters of the stationary DSP are $\alpha$ (optical depth) = 36, $\Omega_{c+} =$ 0.54$\Gamma$ during the input and output stages, $\Omega_{c+} = \Omega_{c-} =$ 0.44$\Gamma$ and $\Omega_a=\Omega_b=0$ during the stationary DSP, $\gamma_{\Lambda}$ (the ground-state decoherence rate) = 9$\times$$10^{-4}$$\Gamma$, and $L\Delta_k$ (the degree of phase mismatch) = 0.90 rad; those of the TPT and DDI are $\Omega_a =$ 2.0$\Gamma$, $\Omega_b = $ 1.6$\Gamma$, $\gamma_R$ (the Rydberg-state decoherence rate) = 0.020$\Gamma$, and $A$ (the DDI coefficient) = 0.60$\Gamma$. (c) The DDI-induced attenuation increased with the Rydberg-state population or equivalently the input probe intensity. The attenuation coefficient (i.e., the logarithm of the ratio of input to output probe energies) without the TPT, $\beta_{\rm DSP}$, and the difference between the attenuation coefficients with and without the TPT, $\Delta\beta$ ($\equiv \beta_{\rm DSP+DDI} - \beta_{\rm DSP}$), as functions of $\Omega_{p0}^2$. Blue and red circles are the data of $\beta_{\rm DSP}$ and $\Delta\beta$, respectively. Based on the above-mentioned parameters, blue and red areas are the predictions with the uncertainties due to the fluctuation of $\pm$1 in $\alpha$ and that of $\pm$1$\times$$10^{-4}$$\Gamma$ in $\gamma_{\Lambda}$.
	}
	\label{fig:SLPTPT}
	\end{figure*}
}
\newcommand{\FigThree}{
	\begin{figure*}[t]
	\includegraphics[width=\textwidth]{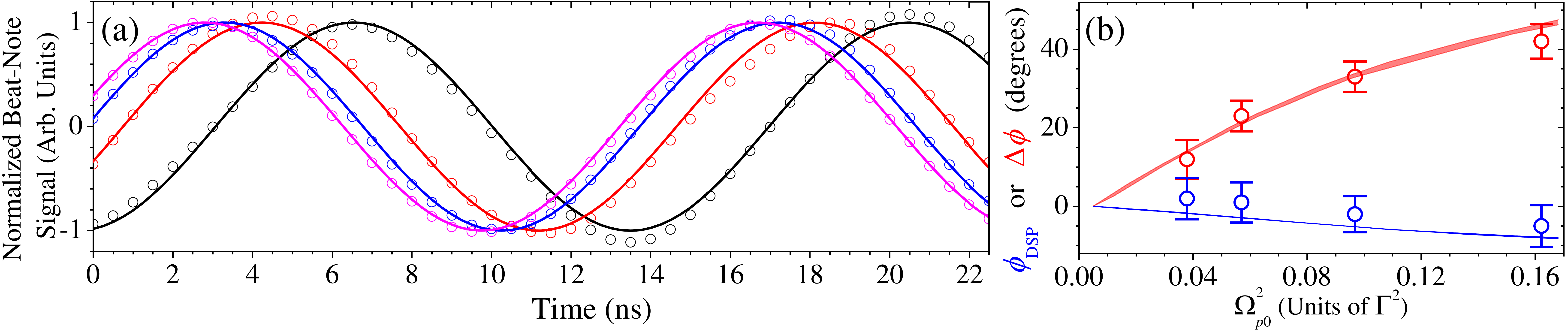}
	\caption{We show quantitatively that the DDI-induced phase shift during the TPT increased with the Rydberg-state population or equivalently the input probe intensity. The analysis of data can be found in text. (a) Representative data of normalized beat-note signals, showing phase evolutions at the peaks of the retrieved probe pulses. Circles are the experimental data and lines are their best fits. In the presence of the TPT (or equivalently the DDI) during the stationary DSP, red, blue, and magenta colors correspond to the peak Rabi frequency of the input probe, $\Omega_{p0}$, of  0.07$\Gamma$, 0.24$\Gamma$, and 0.40$\Gamma$, respectively. The beat-note signals of each $\Omega_{p0}$ in the absence of the TPT were measured, which serve as the reference phases. We make these three signals completely overlap and plot only one here (black color). We make these three signals completely overlap and plot only one here. The experimental parameters are the same as those specified in the caption of Fig.~\ref{fig:SLPTPT}. Since the DDI is negligible at $\Omega_{p0} =$  0.07$\Gamma$, the phase difference between the red and black data is mainly the result of the Rabi oscillation. On the other hand, the phase difference of $+42^\circ$ (or $+23^\circ$) between the magenta (or blue) and red data is the consequence of the DDI. (b) Blue circles are the data of phase shift of the output probe pulse without the TPT, $\phi_{\rm DSP}$, and red circles are those of difference between the phase shifts with and without the TPT, $\Delta\phi$ ($\equiv \phi_{\rm DSP+DDI} - \phi_{\rm DSP}$). The phase shifts at $\Omega_{p0} =$ 0.07$\Gamma$ are subtracted from the data. Blue and red areas are the predictions with the uncertainties due to the fluctuations of $\alpha$ and $\gamma_{\Lambda}$. 
}
	\label{fig:phase}
	\end{figure*}
}
\newcommand{\FigFour}{
\begin{figure}[b]
	\includegraphics[width=\columnwidth]{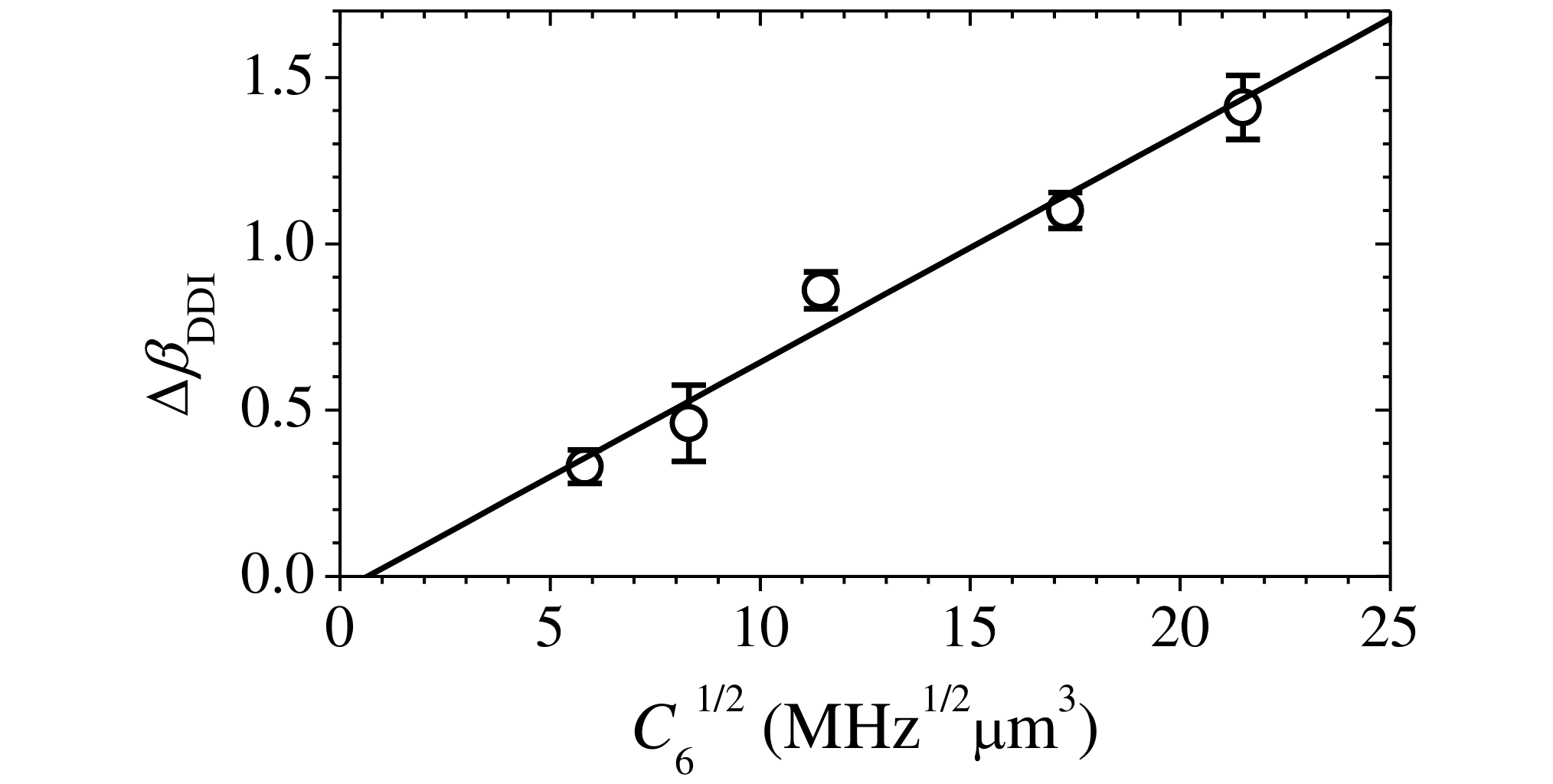}
	\caption{
To further verify that the stationary DSPs carry the DDI, we plot $\Delta\beta_{\rm DDI}$ against $\sqrt{C_6}$, where $\Delta\beta_{\rm DDI}$ is the attenuation coefficient depending only on the DDI strength (see the illustration in text) and $C_6$ is the van der Waals coefficient~\cite{C6}. Circles are the experimental data and straight line is the best fit.
	}
	\label{fig:C6}
	\end{figure}
}

Diluted atomic gases were the first successful physical systems to reach the Bose-Einstein condensation (BEC) by cooling the bosonic atoms below the transition temperatures~\cite{Wieman1995, Ketterle1995}. In such systems, particle-particle interactions are usually weak, or their scattering lengths are typically much less than mean particle spacings---that is the diluteness. The rapid development of optical microcavities makes it possible to realize exciton-polariton BEC in solid-state systems~\cite{Kasprzak2006, BECReview2010, PolaritonThermalTime1, PolaritonThermalTime2, Shishkov2022}. Concerning the uses of Bose condensates, exciton-polariton BECs are limited to two-dimensional systems and have lifetimes comparable to or shorter than thermalization times. 
 
A unique platform of stationary dark-state polaritons (DSPs) to achieve BEC was proposed in Ref.~\cite{Fleischhauer2008}. Compared with the exciton-polariton BEC system, the stationary-DSP BEC system is three-dimensional and has a much longer lifetime. 
The DSPs are bosonic particles and represent the superposition of probe photon and atomic coherence. They are formed by the interaction between a weak probe pulse and atoms under the presence of a strong coupling field based on the effect of electromagnetically induced transparency (EIT) as depicted in Fig.~\ref{fig:transition}(a). The EIT mechanism can store the DSPs in the atoms by turning off the coupling field, and later retrieve the DSPs by turning on the coupling field. Furthermore, when the two counter-propagating coupling fields are applied, the DSPs become stationary and diffuse in the forward and backward directions~\cite{Bajcsy2003, Lin2009, Chen2012, Blatt2016, Campbell2017, KimSLP2018, Everett2019, KimYH2022}.

In Ref.~\cite{Fleischhauer2008}, Fleischhauer {\it et al}. showed that the stationary DSPs are governed by the equation of motion closely similar to the Schr\"{o}dinger equation. They further proposed utilizing a nonlinear Kerr effect to mediate the interaction between the DSPs for thermalization to achieve BEC. However, the proposed Kerr-type interaction is typically too weak to make a sufficient elastic collision rate for thermalization. Therefore, this present work aimed to substitute the dipole-dipole interaction (DDI) between Rydberg-state atoms for the Kerr-type interaction to make the stationary-DSP BEC feasible. 

Rydberg atoms possess strong DDI~\cite{Lukin2001, Tong2004, Heidemann2007, RevModPhys2010, Adams2010}, leading to the applications such as quantum logic gates~\cite{Saffman2005, Keating2015, Tiarks2019, Vaneecloo2022, Stolz2022}, single-photon sources \cite{Ripka2018, Ornelas-Huerta2020, Shi2022}, and strongly-correlated many-body physics~\cite{Pupillo2010, Peyronel2012, Moos2015, Browaeys2020}. In our earlier work, we experimentally demonstrated a many-body system of Rydberg polaritons based on the EIT effect~\cite{OurCommunPhys2021}, where the Rydberg polariton represents the superposition of the photon and the coherence between a Rydberg and a ground state. Slow light arising from the EIT effect greatly enhances the interaction time between light and matter, which can be a couple of $\mu$s to about 10 $\mu$s in a medium of high optical depth (OD)~\cite{OurCommunPhys2021, Chen2012, OurPRL2013, Our2022}. In the thermalization process, the high-OD medium made the interaction time compatible with the elastic collision rate of the $\mu$m$^2$-size interaction cross-section due to the DDI between Rydberg polaritons. Hence, we observed a cooling effect in the transverse direction of slowly-propagating Rydberg polaritons~\cite{OurCommunPhys2021}. 

According to Ref.~\cite{OurCommunPhys2021}, one could create stationary Rydberg polaritons to achieve BEC. The formation of stationary polaritons involves the four-wave mixing (FWM) process \cite{Bajcsy2003, Lin2009, Chen2012, Blatt2016, Campbell2017, KimSLP2018, Everett2019, KimYH2022}. The ladder-type transition scheme, which typically has a very large phase mismatch in the FWM process, is employed in the Rydberg-EIT system to form the stationary Rydberg polaritons. However, the probe and coupling fields had the typical wavelengths of 780 or 795 nm and around 480 nm in a Rydberg-EIT system, resulting in an FWM phase mismatch of $10^4$--$10^5$~rad. Such a large phase mismatch completely destroys stationary Rydberg polaritons and makes the search for the Rydberg-polariton BEC impractical.

\FigOne

In this study, we proposed and experimentally demonstrated the stationary DSPs possessing the Rydberg-state DDI. In addition to the probe and coupling fields that formed the stationary DSP via the FWM process, two more laser fields were applied to drive the two-photon transition (TPT) of $|2\rangle \rightarrow |4\rangle \rightarrow |5\rangle$ as shown by Fig.~\ref{fig:transition}(a). The TPT generated the Rabi oscillation between the population of the ground state $|2\rangle$ and that of the Rydberg state $|5\rangle$, as well as between the ground-state coherence $\rho_{21}$ and the Rydberg coherence $\rho_{51}$. Due to the existence of the Rydberg population, the DDI resulted in the phase shift and attenuation of the stationary DSP. 
With regard to the thermalization of stationary DSPs, the DDI can lead to a far larger elastic collision rate than the Kerr-type interaction. Hence, this work makes a substantial advancement toward the realization of the stationary-DSP BEC.


We carried out the experiment in laser-cooled $^{87}$Rb atoms with a temperature of about 350 $\mu$K. Before each measurement, the magnetic and laser fields for the production of the cold atoms were switched off, and we optically pumped all the population to a single Zeeman state. Details of the atom cloud and the experimental procedure before the measurements can be found in Refs.~\cite{OurPRA2019, OurCommunPhys2021}. 

All the laser fields had the $\sigma_+$ polarization in the experiment. In the $\Lambda$-type EIT system shown by Fig.~\ref{fig:transition}(a), $|1\rangle$, $|2\rangle$, and $|3\rangle$ are $|5S_{1/2}, F=1, m_F=1\rangle$, $|5S_{1/2}, F=2, m_F=1\rangle$, and $|5P_{3/2}, F=2, m_F=2\rangle$. The EIT was driven resonantly, while the fields in the forward direction ($\Omega_{p+}$ and $\Omega_{c+}$) had nearly zero one-photon detuning, and those in the backward direction ($\Omega_{p-}$ and $\Omega_{c-}$) had the one-photon detuning $\Delta_-$ of $-1$$\Gamma$~\cite{Lin2009}. In the TPT system shown by Fig.~\ref{fig:transition}(a), $|4\rangle$ is $|5P_{3/2}, F=3, m_F=2\rangle$ and $|5\rangle$ is $|32D_{5/2}, m_J=3/2~\&~5/2\rangle$. We made the TPT resonant at nearly no DDI and set the one-photon detuning $\Delta_a$ to +5$\Gamma$ to make excitation to $|4\rangle$ negligible. 

The propagation direction of $\Omega_{c-}$ was exactly opposite to that of $\Omega_{c+}$ as depicted in Fig.~\ref{fig:transition}(b). When interacting with the atoms, $\Omega_{c+}$ and $\Omega_{p+}$ propagated in the nearly same direction with an angle separation of about 0.3$^{\circ}$~\cite{Lin2009}. The backward probe field $\Omega_{p-}$, or more precisely $\Omega_{p-}(z,t)$, depicted in Fig.~\ref{fig:transition}(a) only appeared during the stationary DSP, and its value was nearly the same as $\Omega_{p+}(z,t)$~\cite{Lin2009}. Driving the transitions of $2\rangle$ $\rightarrow$ $|4\rangle$ and $|4\rangle$ $\rightarrow$ $|5\rangle$, the TPT fields with Rabi frequencies of $\Omega_a$ and $\Omega_b$ counterpropagated in the forward and backward directions as depicted in Fig.~\ref{fig:transition}(c). The laser beams of $\Omega_a$ and $\Omega_b$ completely covered the region of stationary DSPs. Other details of the experimental setup can be found in \SMexp.

\FigTwo

We made theoretical predictions using the optical Bloch equations of the density-matrix operator and the Maxwell-Schr\"{o}dinger equations of the probe fields. Details of the equations and calculation can be found in \SMeq. In our earlier works~\cite{OurPRR2022,OurOEMFT}, the predictions are in good agreement with the experimental data.

We set $\delta_\Lambda = 0$ in the experiment, where $\delta_\Lambda$ is the two-photon detuning in the $\Lambda$-type EIT system. In Ref.~\cite{Tebbenv2021}, Tebben {\it et al}.~theoretically studied a similar transition scheme, except that the TPT is replaced by a one-photon transition. They showed that the optimum $\delta_\Lambda$, which maximizes the stationary-DSP energy, is equal to a half of the Rabi frequency of the one-photon transition~\cite{OurIDR}. In our TPT case, $\delta_\Lambda = 0$ is the optimum value.

Representative data that demonstrate the formation of stationary DSPs in the $\Lambda$-type EIT system are shown by Fig.~\ref{fig:SLPTPT}(a) (\ref{fig:SLPTPT}(b)) without (with) the TPT. The Rabi oscillation between the population in $|2\rangle$ and that in $|5\rangle$ (or between coherences $\rho_{21}$ and $\rho_{51}$) was clearly observed during the magenta pulse in Fig.~\ref{fig:SLPTPT}(b). The consistency between the data and the predictions is satisfactory. More details of the data are described in Sec.~III of the SM.

In the measurements of the DSPs without and with the TPT such as Fig.~\ref{fig:SLPTPT}, the Rabi frequencies $\Omega_{c+}$, $\Omega_{c-}$, and $\Omega_a$, the optical density or abbreviated as OD ($\alpha$), and the ground-state decoherence rate ($\gamma_{\Lambda}$) were pre-determined \cite{OurPRA2019, PhotonSwitching, OurPRL2006}, where the probe transmission without the EIT is indicated by $\exp(-\alpha)$ and the decay rate of the coherence $\rho_{21}$ is represented by $\gamma_{\Lambda}$. Details of the determination methods and representative data are presented in \SMexpANDrb. The Rabi frequency $\Omega_b$ was determined by the period of the Rabi oscillation during the TPT. The stationary DSP is generated by the FWM process, in which the phase mismatch causes the energy loss~\cite{Chen2012, Ldk2004, FBS2021}. The degree of phase mismatch is given by $L\Delta_k$ [see \SMldk]. We determined the value of $L\Delta_k$ by the comparison between the experimental results and theoretical predictions. Details can be found in \SMslp, which provides more evidence for the formation of stationary DSPs.

\FigThree

We varied the input probe power~\cite{LargeOmega_p0}, while keeping the pulse width and beam profile the same, and measured the data similar to those in Figs.~\ref{fig:SLPTPT}(a) and \ref{fig:SLPTPT}(b). The population in $|2\rangle$, $\rho_{22}$, is about equal to $|\Omega_{p+}/\Omega_{c+}|^2$ due to the EIT effect. Thus, a larger input probe intensity, $\Omega_{p0}^2$, resulted in larger $\rho_{22}$ and $\rho_{55}$, which produced a higher DDI strength \cite{OurCommunPhys2021}. We determined the attenuation coefficients, $\beta_{\rm DSP}$ and $\beta_{\rm DSP+DDI}$, as functions of $\Omega_{p0}^2$, where $\beta_{\rm DSP}$ and $\beta_{\rm DSP+DDI}$ are defined as the logarithm of the ratio of input to output probe energies without and with the DDI, and $\Omega_{p0}^2$ is the square of the Rabi frequency of the input probe pulse peak. In Fig.~\ref{fig:SLPTPT}(c), the blue and red circles represent the data of $\beta_{\rm DSP}$ and $\Delta\beta$ ($\equiv \beta_{\rm DSP+DDI} - \beta_{\rm DSP}$). Since the OD fluctuated about $\pm$1 and the ground-state decoherence rate $\gamma_{\Lambda}$ fluctuated about $\pm1$$\times$$10^{-4}$$\Gamma$, the predictions of $\beta_{\rm DSP}$ and $\Delta\beta$ are plotted as the blue and red areas. 

The TPT made the population (coherence) oscillate between $|2\rangle$ and $|5\rangle$ (between $\rho_{21}$ and $\rho_{51}$), and the population in $|5\rangle$ induced the DDI. We characterized the DDI coefficient, $A$, which is defined as the decoherence rate (also frequency shift) per $\rho_{55}$~\cite{OurOEMFT, OurCommunPhys2021, OurPRR2022}. See also \SMddiA~for the definition of $A$. The ratio of retrieved probe energies with to without the TPT was measured against the input probe intensity or equivalently the peak Rabi frequency square, $\Omega_{p0}^2$, as shown in Fig.~\ref{fig:SLPTPT}(c). We compared the data with the predictions to determine $A =$ 0.60$\Gamma$ and $\gamma_R$ (the decay rate of the coherence $\rho_{51}$)  = 0.020$\Gamma$. We also estimated the value of $A$, and details can be found in \SMddi. The estimation gives $A$ of 0.54$\Gamma$, indicating that the experimentally determined $A$ of 0.60$\Gamma$ is reasonable.

The DDI induces a phase shift of the stationary DSP, and the phase of the retrieved probe pulse is shifted. To further verify the creation of the stationary DSP dressed by the DDI, we applied the TPT and measured the phase shift of the retrieved probe pulse, i.e., the difference of the phases with and without the atoms. The beat-note interferometer was employed to measure the phase evolution around the pulse peak~\cite{beatnote, OurCommunPhys2021}. The experimental parameters in the phase measurement are the same as those in the transmission measurement. In Fig.~\ref{fig:phase}(a), the black circles are the beat-note data without the TPT or DDI, which serve as the reference for the other data. The red, blue, and magenta circles represent the data with increasing values of $\Omega_{p0}^2$ or DDI strength. A larger DDI strength resulted in a larger phase shift as expected. 

We measured the phase shifts of the retrieved probe pulses without and with the TPT, $\phi_{\rm DSP}$ and $\phi_{\rm DSP+DDI}$, respectively, as functions of $\Omega_{p0}^2$. In Fig.~\ref{fig:phase}(b), the blue and red circles represent $\phi_{\rm DSP}$  and $\Delta\phi$ ($\equiv \phi_{\rm DSP+DDI} - \phi_{\rm DSP}$). We subtracted the measured phase shift at $\Omega_{p0} =$ 0.07$\Gamma$ from the data. The subtraction removes the phase shift contributed from the Rabi oscillation. Thus, $\Delta\phi$ exhibits mainly the DDI effect. More details can be found in \SMphase. Without the DDI, $\phi_{\rm DSP}$ of the stationary DSP depends on the probe intensity a little. Owing to the DDI, $\Delta\phi$ depends on the probe intensity significantly. The blue and red areas are the theoretical predictions. In the theoretical calculation, all the parameters are non-adjustable and experimentally pre-determined. The consistency between the data and predictions is satisfactory, confirming that the stationary DSP indeed possessed the DDI.

\FigFour

As another evidence of the stationary DSPs carrying the DDI, we measured the DDI effect at different Rydberg states $|nD_{5/2}\rangle$ with the principal quantum numbers, $n$, of 28, 30, 32, 35, and 38. 
The DDI potential energy between two Rydberg atoms is given by $C_6/r^6$, where $C_6$ is van der Waals coefficient and $r$ is the distance between the atoms. The DDI energy shift or DDI-induced attenuation coefficient is linearly proportional to $\sqrt{C_6}$~\cite{OurOEMFT,SqrtC6-1, SqrtC6-2, SqrtC6-3}. In Fig.~\ref{fig:C6}, we plot the experimental data of DDI-induced attenuation coefficient, $\Delta\beta_{\rm DDI}$, against $\sqrt{C_6}$, where $\Delta\beta_{\rm DDI}$ is the difference between the DDI-induced attenuation coefficient of $\Omega_{p0} = 0.24$$\Gamma$ and that of $\Omega_{p0} = 0.07$$\Gamma$. This $\Delta\beta_{\rm DDI}$ can avoid any attenuation effect that depends on $n$, e.g., different two-photon frequency fluctuations due to different $n$'s, except the DDI effect.
The values of $C_6$ were obtained from the programming code provided by Ref.~\cite{C6}. The experimental data of $\Delta\beta_{\rm DDI}$ clearly exhibit the linear dependence on $\sqrt{C_6}$, further confirming that the stationary DSPs possess the DDI.

We now estimate whether the present experimental condition is close to the observation of stationary-DSP BEC. Details of the estimations can be found in \SMbec. Based on the formulas in Ref.~\cite{Fleischhauer2008}, the BEC transition temperature, $T_c$, is about 4.0~mK and the stationary-DSP temperature, $T_p$, is around 3.8 $\mu$K. The measured phase shift indicates that the elastic collision rate, $R_c$, is approximately 33 $\mu$s$^{-1}$. Under $T_p \ll T_c$, such $R_c$ enables the thermal equilibrium of stationary DSPs, and makes the BEC feasible. To observe the BEC, we still need to build an artificial trap and produce stationary DSPs in the quasi-continuous mode.

In conclusion, we experimentally demonstrated the formation of the stationary DSP dressed by the DDI, using the scheme of the $\Lambda$-type EIT system together with the TPT-driven Rabi oscillation between a ground state and a Rydberg state. The scheme overcomes the severe problem of a large phase mismatch in the direct formation of the stationary Rydberg polariton. As proposed in Ref.~\cite{Fleischhauer2008}, the system of stationary DSPs is the possible platform for a three-dimensional and long-lifetime Bose condensate. Our work made the stationary DSPs carry the DDI and provided a feasible method of thermalization for the realization of BEC.


\section*{ACKNOWLEDGMENTS}

This work was supported by Grants No.~110-2639-M-007-001-ASP and No.~111-2639-M-007-001-ASP of the National Science and Technology Council, Taiwan, and Grant No.~111-2923-M-008-004-MY3 of the Mutual Funds for Scientific Cooperation between Taiwan, Latvia, and Lithuania. The authors thank the fruitful discussions with Prof. Gediminas Juzeli\={u}nas and Mr. Chin-Jen Yang.


\end{document}


\title{
SUPPLEMENTAL MATERIAL \\
\vspace*{\baselineskip}
Experimental Demonstration of Stationary Dark-State Polaritons Dressed by Dipole-Dipole Interaction
}

\author{
Bongjune Kim,$^{1}$
Ko-Tang Chen,$^{1}$ 
Kuei-You Chen,$^{1}$ 
Yu-Shan Chiu,$^{1}$
Chia-Yu Hsu,$^{1}$
Yi-Hsin Chen,$^{2,3}$ 
and 
Ite A. Yu$^{1,3}$}

\affiliation{
$^{1}$Department of Physics, National Tsing Hua University, Hsinchu 30013, Taiwan \\
$^{2}$Department of Physics, National Sun Yat-sen University, Kaohsiung 80424, Taiwan \\
$^{3}$Center for Quantum Science and Technology, National Tsing Hua University, Hsinchu 30013, Taiwan
}

\maketitle

\newcommand{\FigSOne}{
\begin{figure*}[t]
	\includegraphics[width=0.9\textwidth]{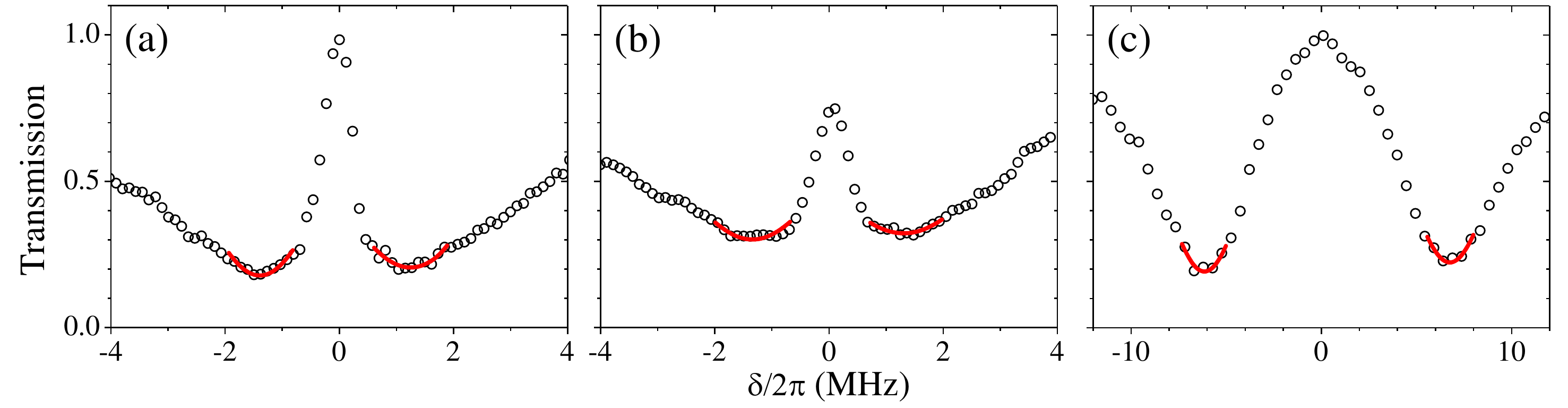}
	\caption{(a) [or (b)] Probe transmission spectrum as a function of the two-photon detuning, $\delta$, under the constant presence of the forward (or backward) coupling field of $\Omega_{c+}$ (or $\Omega_{c-}$) in the $\Lambda$-type EIT system, and (c) that under the constant presence the TPT field of $\Omega_b$ in the Rydberg-EIT system. In the measurement of each spectrum, the one-photon detuning was nearly zero. Black circles are the experimental data. Red curves are the best fits to determine the frequencies of minima. The frequency separation between the two minima gives the Autler-Townes splitting of (a) 0.44$\Gamma$, (b) 0.44$\Gamma$, or (c) 2.0$\Gamma$.
}
	\label{fig:ATS}
	\end{figure*}
}
\newcommand{\FigSTwo}{
\begin{figure}[b]
	\includegraphics[width=0.9\columnwidth]{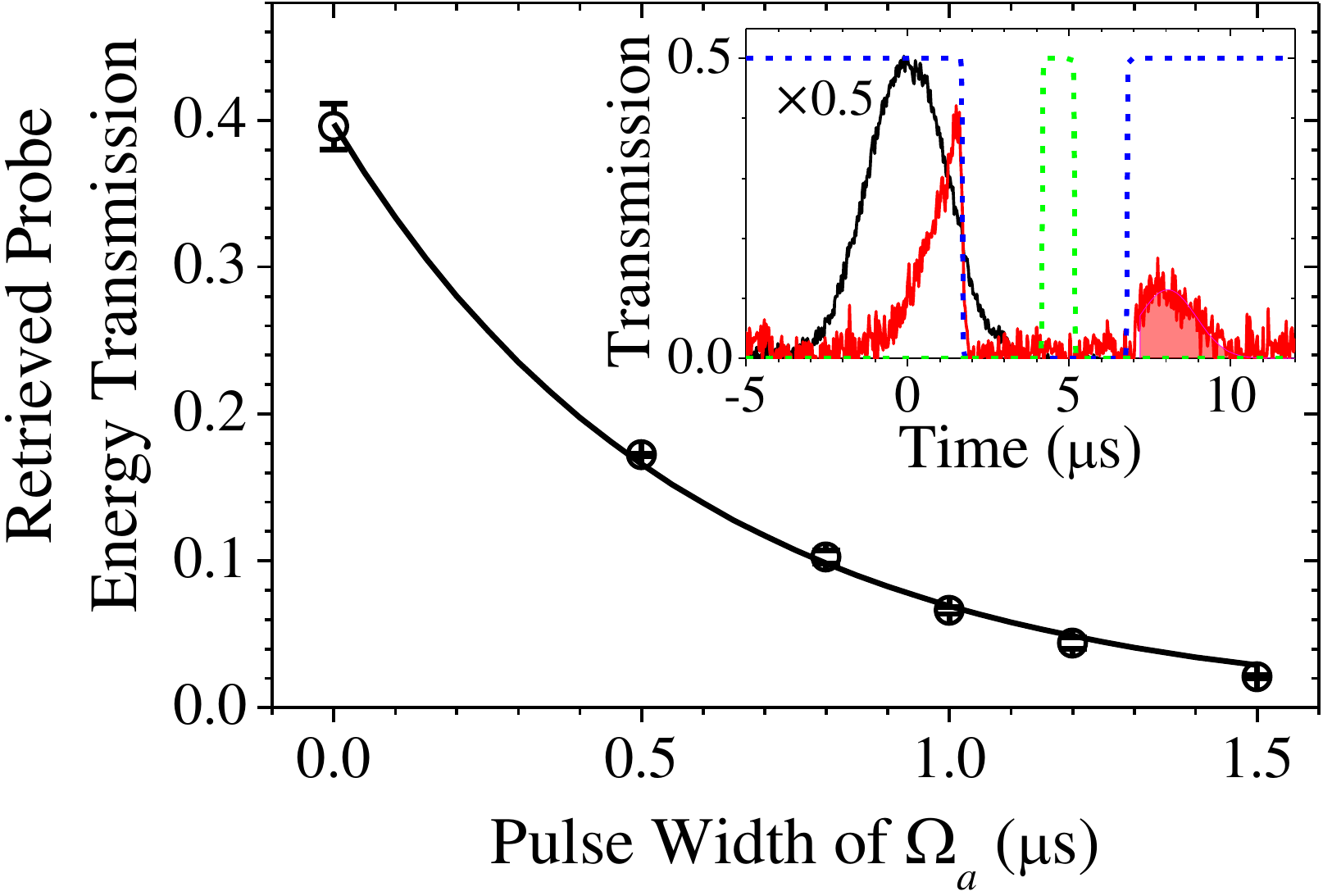}
	\caption{Main plot: In the system of all-optical switching (AOS), transmission of the retrieved probe energy versus the pulse duration of the field of $\Omega_a$. Circles are the experimental data taken at $\Delta_a = $ 5.0$\Gamma$. Black line is the best fit with the fitting function given by Eq.~(\ref{eq:AOS}), which determines $\Omega_a =$ 2.0$\Gamma$. Inset: Black and red lines are the signals of the probe pulses without and with the atoms, respectively. The black line is scaled down by a factor of 0.5. Blue and green dashed lines represent the time sequences of the coupling field and the field of $\Omega_a$. Filled area in red represents the retrieved probe energy after the AOS effect.
}
	\label{fig:AOS}
	\end{figure}
}
\newcommand{\FigSThree}{
\begin{figure}[t]
	\includegraphics[width=0.9\columnwidth]{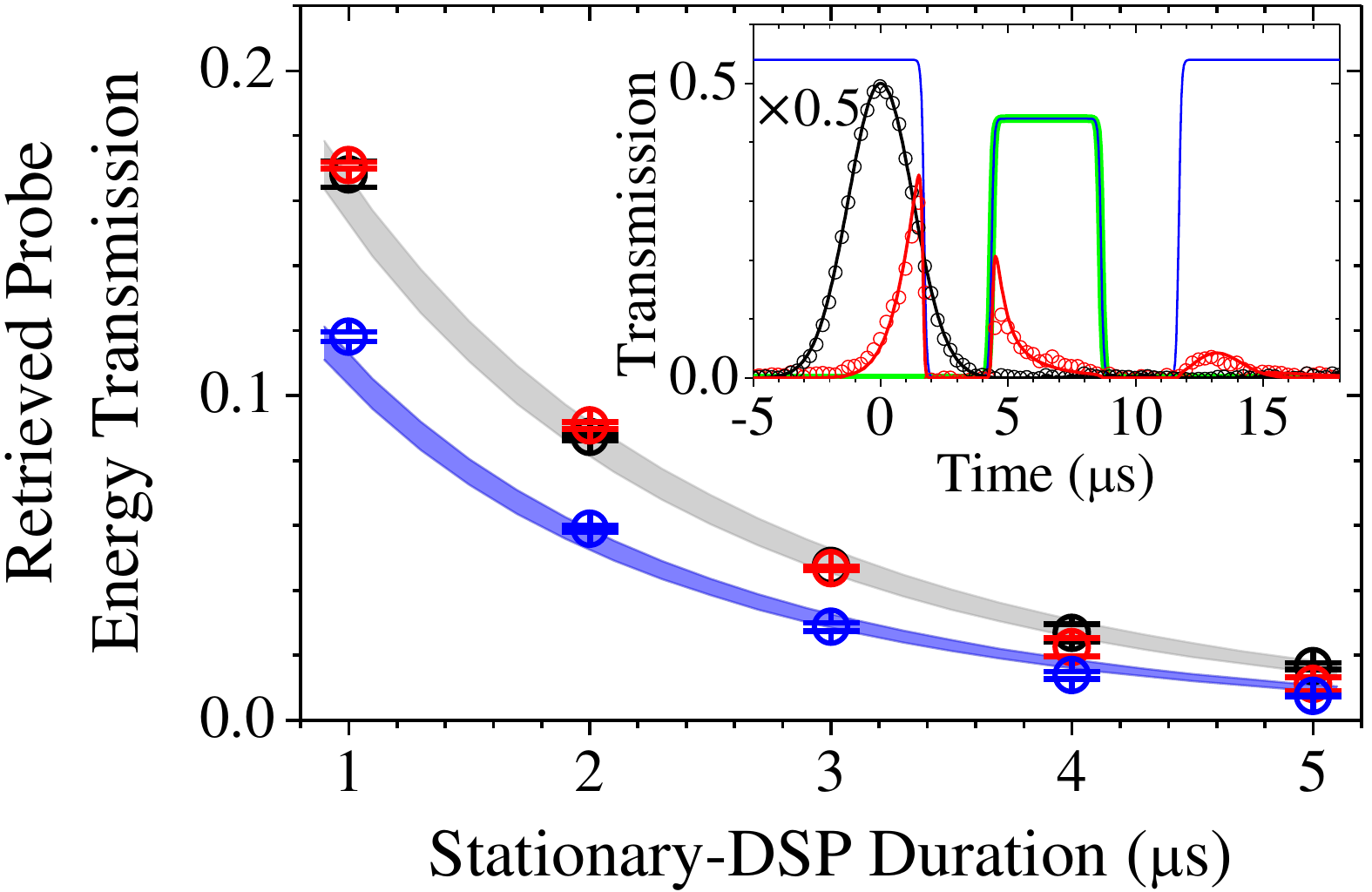}
	\caption{Main plot: Transmission of the retrieved probe energy versus the stationary-DSP duration. Black, red, and blue circles are the experimental data taken at $\Omega_{p0} =$ 0.07$\Gamma$, 0.24$\Gamma$, and 0.40$\Gamma$, respectively, where $\Omega_{p0}$ is the peak Rabi frequency of the input probe pulse. Gray ($\Omega_{p0} =$ 0.07$\Gamma$ and 0.24$\Gamma$) and blue ($\Omega_{p0} =$ 0.40$\Gamma$) areas are the theoretical predictions calculated with $L\Delta_k$  = 0.90 rad. The experimental parameters were $\alpha$ = 36$\pm$1, $\Omega_{c+} =$ 0.54$\Gamma$ before the storage and after the retrieval, $\Omega_{c+} = \Omega_{c-} =$ 0.44$\Gamma$ during the stationary DSP, and
$\gamma_{\Lambda}$ = (9$\pm$1)$\times$$10^{-4}$$\Gamma$. Thicknesses of the gray and blue areas represent the uncertainties or fluctuations in the OD and ground-state decoherence rate. Inset: Representative data of the probe signals without (black circles) and with (red circles) the atom cloud. Black (red) line is the best fit (theoretical prediction) of the black (red) circles. The black circles and line are scaled down by a factor of 0.5. Blue and green lines represent the timing sequence of $\Omega_{c+}$ and $\Omega_{c-}$, the two lines completely overlap during the stationary DSP (4 $\mu$s $< t <$ 9 $\mu$s here). The red circles underneath the green-line pulse reveal that during the stationary DSP the probe light leaked out of the atoms in the forward direction.
}
	\label{fig:StationaryDSP}
	\end{figure}
}
\newcommand{\FigSFour}{
\begin{figure}[b]
	\includegraphics[width=0.89\columnwidth]{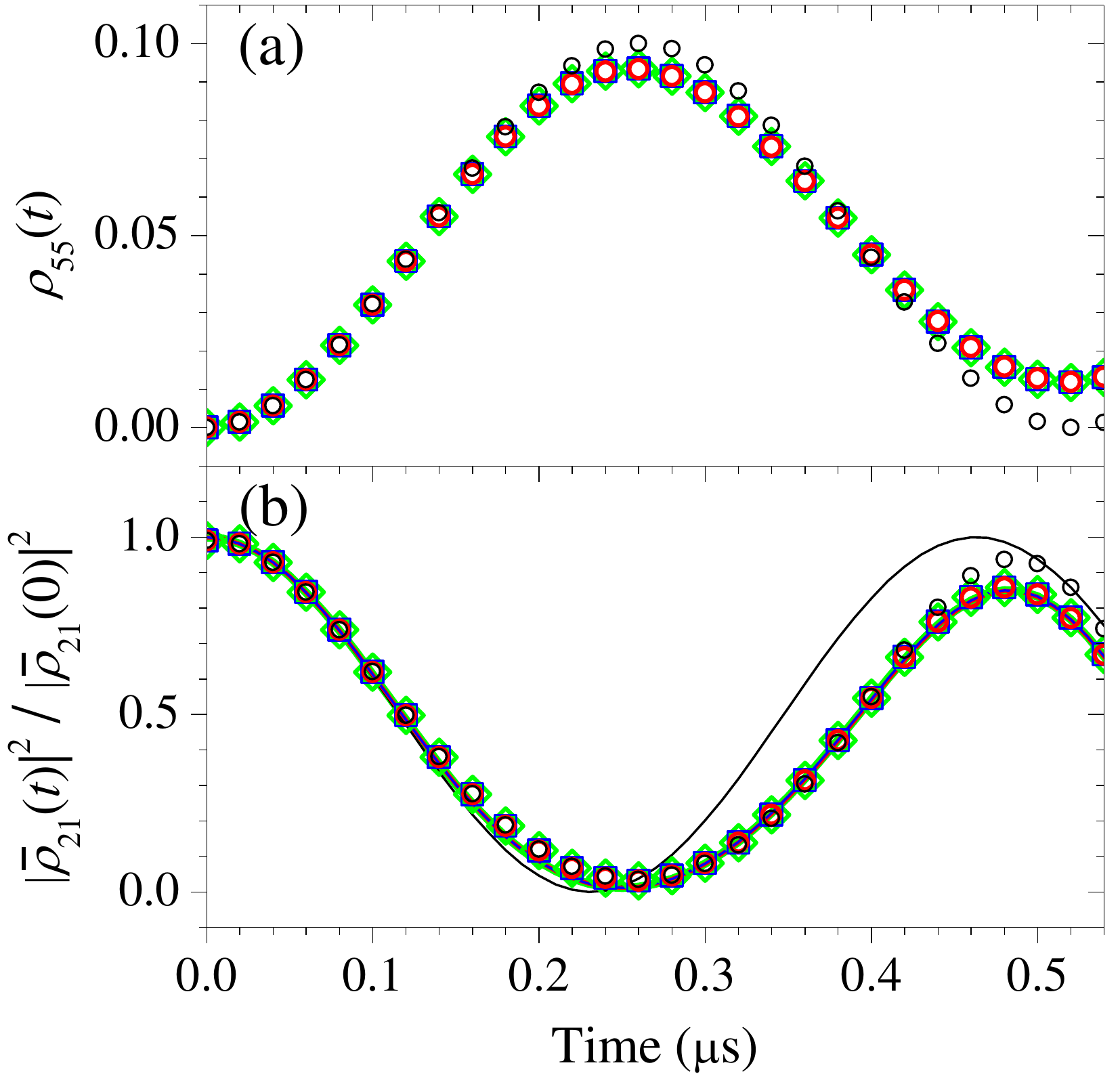}
	\caption{Estimation of the DDI strength or coefficient $A$. (a) The solution of $\rho_{55}(t)$ by numerically solving Eqs.~(\ref{eq:OBEopt1})-(\ref{eq:OBEopt3}). (b) The ensemble average of the microscopic DDI effect, i.e., $|\bar{\rho}_{21}|^2$ calculated from Eqs.~(\ref{eq:bar_rho21r})-(\ref{eq:bar_rho21sq}). Black circles, red circles, blue squares, and green diamonds represent the results from the first to the fourth iterations. Lines represent the corresponding best fits with the fitting function of $|\rho_{21}(\omega =A\rho_{55}, \gamma = A\rho_{55})|^2$ given by Eq.~(\ref{eq:rho21sq}), in which $A$ is the fitting parameter.
}
	\label{fig:EstimateA}
	\end{figure}
}
\newcommand{\FigSFive}{
\begin{figure}[t]
	\includegraphics[width=0.9\columnwidth]{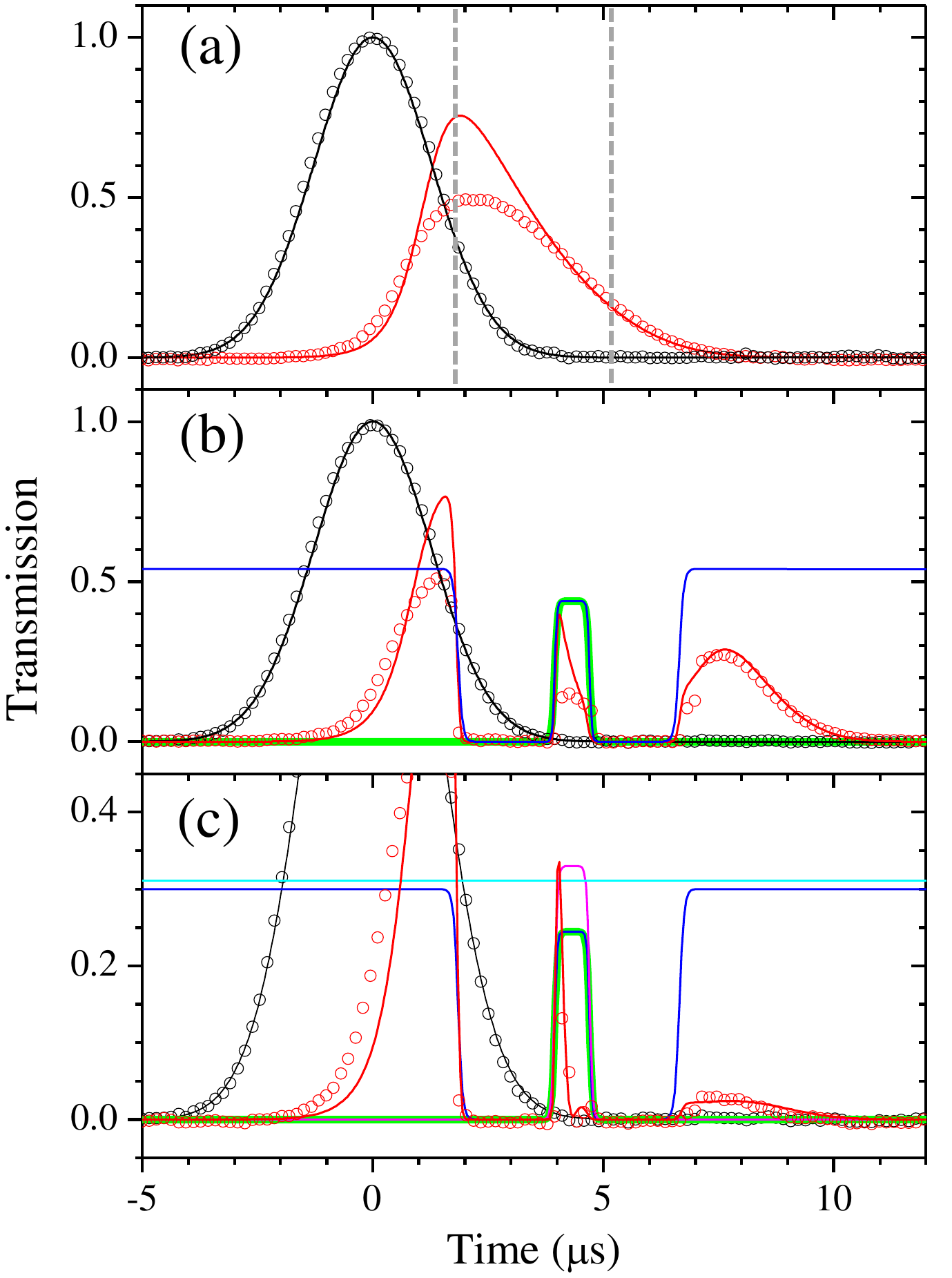}
	\caption{(a) Slow light under the constant presence of the coupling field. (b, c) The formation of the stationary DSP without and with the TPT or equivalently DDI, respectively. $\Omega_{p0} =$ 0.4$\Gamma$, and all the other experimental parameters are the same as those shown in the caption of \MainFigTwo. Black and red circles are the experimental data of the input and output probe pulses. Black lines represent the Gaussian best fits and red lines are the theoretical predictions. Blue, green, magenta, and cyan lines show the timing sequences of the coupling fields of $\Omega_{c+}$ and $\Omega_{c-}$ as well as the TPT fields of $\Omega_a$ and $\Omega_b$. The part of the probe pulse between the two vertical gray dashed lines in (a) was stored in the atoms at $t \approx$ 6.0 $\mu$s (when $\Omega_{c+}$ was switched off) in (b) and (c).
}
	\label{fig:LargeProbe}
	\end{figure}
}
\newcommand{\FigSSix}{
\begin{figure}[t]
   \includegraphics[width=0.9\columnwidth]{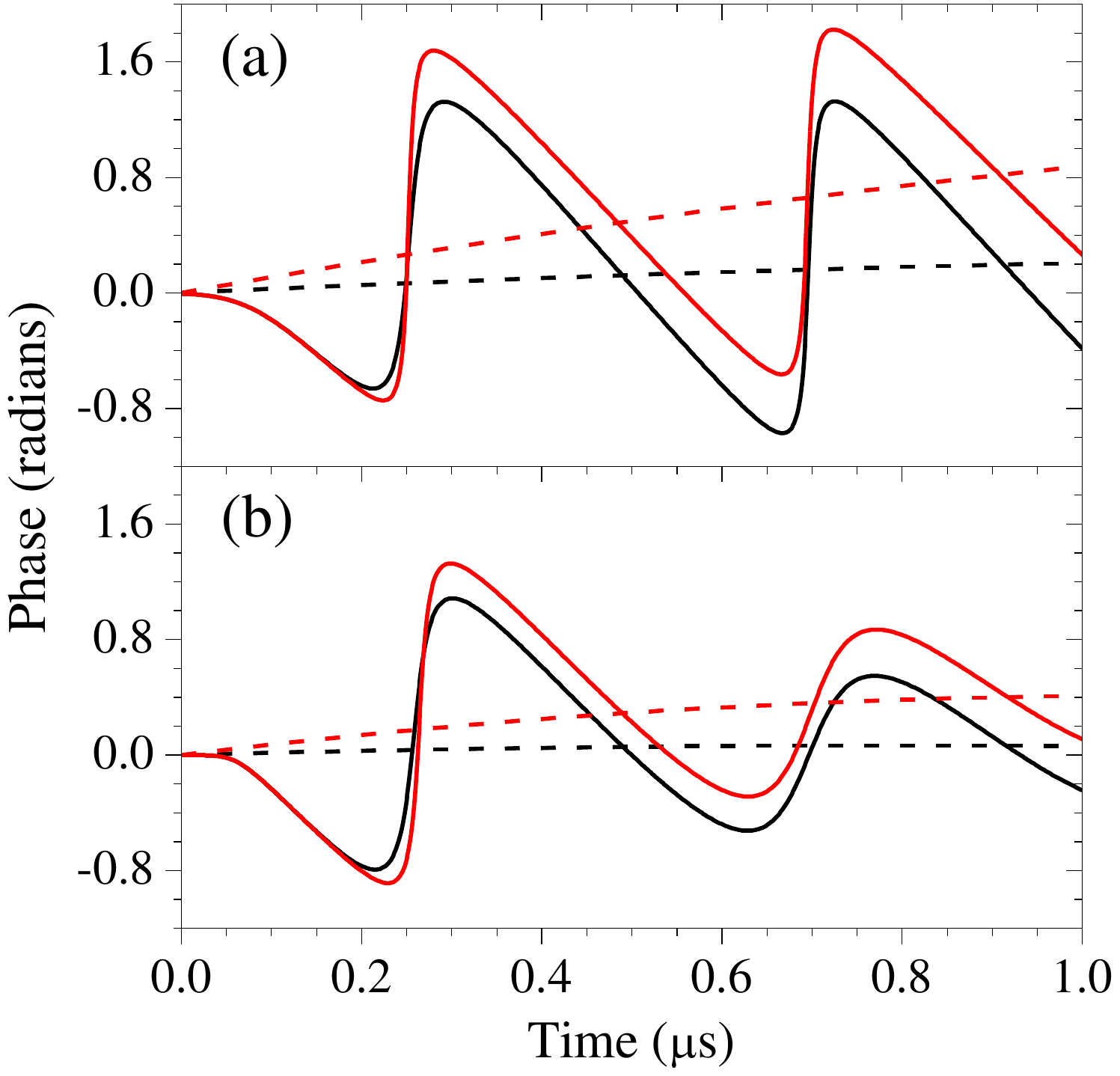}
    \caption{Simulations of the phase shift of the ground-state coherence $\rho_{21}$, or  equivalently the phase shift of the output probe pulse, as a function of time. Black and red lines are the results without and with the DDI of $A =$ 0.60$\Gamma$, respectively. In the simulations, $\Omega_{p0} =$ 0.24$\Gamma$, $\Delta_- =0$, $\Omega_a = \Omega_b =$ 2.0$\Gamma$, and the remaining parameters are the same as those in the caption of \MainFigTwo. All decay and decoherence mechanisms are switched off in (a) and intact in (b). Dashed lines represent the baselines of the oscillations. In each subfigure, the difference between the two dashed lines reveals the DDI-induced phase shift.
       }
    \label{fig:PhaseShift}
\end{figure}
}
\section{Experimental Setup}
\label{sec:ExpSetup}

In the system for the creation of the stationary dark-state polaritons (DSP)~\cite{SLP}, there are two ground states $|1\rangle$ ($|5S_{1/2},F$=$1,m_F$=$1\rangle$) and $|2\rangle$ ($|5S_{1/2},F$=$2,m_F$=$1\rangle$) and an excited state $|3\rangle$ ($|5P_{3/2},F$=$2,m_F$=$2\rangle$) to form the $\Lambda$-type configuration of the electromagnetically induced transparency (EIT) effect. Since laser-cooled $^{87}$Rb atoms were employed in the experiment, all the energy levels are referred to those of $^{87}$Rb. The coupling (or probe) fields propagated in the forward and backward directions, and their Rabi frequencies are denoted as $\Omega_{c+}$ and $\Omega_{c-}$ (or $\Omega_{p+}$ and $\Omega_{p-}$), respectively. Both coupling fields drove the transition from $|2\rangle$ to $|3\rangle$, and both probe fields drove that from $|1\rangle$ to $|3\rangle$. All the fields had the $\sigma+$ polarization. The peak Rabi frequency of the input Gaussian probe pulse (denoted as $\Omega_{p0}$) ranged from $0.07$$\Gamma$ to $0.40$$\Gamma$, where $\Gamma$ ($= 2$$\pi$$\times$6 MHz) is the spontaneous decay rate of $|3\rangle$. Throughout this work, $\Omega_{c+} =$ 0.54$\Gamma$ during the input and output stages of the probe pulse, and $\Omega_{c+} = \Omega_{c-} =$ 0.44$\Gamma$ during the stationary DSP. The determination method of $\Omega_{c+}$ and $\Omega_{c-}$ will be described in \SecRabiFreq.

The lasers that generated the coupling and probe fields were injection-locked to a master laser. Because of the injection lock, the contribution of the laser frequency fluctuations to the decoherence rate in the $\Lambda$-type EIT system is negligible~\cite{OurPRA2019}. In the forward direction, we maintained the frequencies of $\Omega_{p+}$ and $\Omega_{c+}$ around the resonance frequencies of their transitions ($\Delta_{+} \approx 0$), and kept their frequency difference, i.e., their two-photon frequency, to the two-photon resonance. In the backward direction, we also kept the two-photon frequency of $\Omega_{p-}$ and $\Omega_{c-}$ to the two-photon resonance, but red-detuned their frequencies by about 1.0$\Gamma$ with respect to their transition frequencies ($\Delta_{-} =$ $-1.0$$\Gamma$). A sufficiently large difference between the one-photon detuning in the forward direction, i.e., $\Delta_{+}$, and that in the backward direction, i.e., $\Delta_{-}$, is necessary for the formation of stationary DSPs in cold media~\cite{SLP}. 

In the system of the two-photon transition (TPT) that induced the dipole-dipole interaction (DDI), there are a ground state $|2\rangle$ which is the same state $|2\rangle$ in the $\Lambda$-type EIT system, an excited state $|4\rangle$ ($|5P_{3/2},F$=$3,m_F$=$2\rangle$), and a Rydberg state $|5\rangle$ ($|32D_{5/2},m_J$=$3/2\,\&\,5/2\rangle$). The spontaneous decay rate of $|4\rangle$ is equal to $\Gamma$, and that of $|5\rangle$ is 2$\pi$$\times$7.9~kHz or 1.3$\times$$10^{-3}$$\Gamma$. The two laser fields drove the transition from $|2\rangle$ to $|4\rangle$ and that from $|4\rangle$ to $|5\rangle$, and they had the Rabi frequencies $\Omega_{a}$ and $\Omega_{b}$, respectively. Throughout this work, $\Omega_{a} =$ 2.0$\Gamma$ and $\Omega_{b} =$ 1.6$\Gamma$ during the TPT. The determination method of $\Omega_{a}$ and $\Omega_{b}$ will be described in \SecRabiFreq.

The frequencies of the TPT fields of $\Omega_{a}$ and $\Omega_{b}$ were stabilized with the Pound-Drever-Hall scheme. The root-mean-square value of the two-photon frequency fluctuation is about 150~kHz. We utilized acousto-optic modulators (AOMs) and electro-optic modulators (EOMs) to tune frequencies and to modulate amplitudes of the laser fields. The rf or microwave frequency sources of the AOMs and EOMs were locked to a rubidium frequency standard (SRS FS725). Details of the stabilization and tuning of the laser field frequencies can be found in Ref.~\cite{OurPRA2019}. We fixed the sum of the frequencies, i.e., two-photon frequency, of $\Omega_{a}$ and $\Omega_{b}$ to the two-photon resonance, while detuning their frequencies by 5.0$\Gamma$ with respect to their transition frequencies ($\Delta_{a} = -\Delta_{b} =$ +5.0$\Gamma$).

In the $\Lambda$-type EIT system, the $e^{-1}$ full width of the input probe beam was 130~$\mu$m, and those of the forward and backward coupling beams were 3.1~mm. The forward and backward coupling fields completely overlapped and covered the entire atom cloud. 

In the TPT system, the field of $\Omega_{a}$ propagated in the forward direction and that of $\Omega_{b}$ in the backward direction. This counter-propagation configuration minimized the Doppler effect in the TPT. The $e^{-1}$ full width of the $\Omega_a$ beam was 4.9~mm, and that of the $\Omega_b$ beam was 250~$\mu$m. The beams of $\Omega_{a}$ and $\Omega_{b}$ fully covered the region of stationary DSPs in
the atom cloud.

The output probe field was coupled to a single-mode optical fiber to minimize the background signals from the other light fields, and detected by a photomultiplier tube (PMT). A digital oscilloscope (Agilent MSO6014A) acquired the signal from the PMT. All the timings of $\Omega_{c+}^2$, $\Omega_{c-}^2$, $\Omega_{a}^2$, and $\Omega_{b}^2$ were monitored by fast photo detectors. In the measurement of the phase shift of the output probe pulse, we employed the beat-note interferometer~\cite{beatnote}. The measurement method of the phase shift was the same as that described in Ref.~\cite{OurCommunPhys2021}. 

In the experiment, the optical depth (OD) of the system was maintained around 36. Using a known value of $\Omega_{c+}$ in the $\Lambda$-type EIT system, we measured the propagation delay time, $\tau_d$, of the slow-light probe pulse to determine the OD, $\alpha$, where $\tau_d = \alpha\Gamma/\Omega_{c+}^2$~\cite{OurPRA2019}. The ground-state decoherence rate, $\gamma_{\Lambda}$, affects the coherence involving the DSP, and its value was about 9$\times$$10^{-4}$$\Gamma$ or 2$\pi$$\times$5.4~kHz in this work. We measured the output transmission of the probe field, $T_{\rm SL}$, to determine the value of $\gamma_{\Lambda}$, where $T_{\rm SL} = \exp(-2\alpha\gamma_{\Lambda}\Gamma/\Omega_{c+}^2)$~\cite{OurPRA2019}. Examples of data for the determinations of the coupling Rabi frequency, optical density, and decoherence rate in a $\Lambda$-type EIT system can be found in Fig.~3 of Ref.~\cite{OurPRA2019}.

\section{Theoretical Model}
\label{sec:Theory}

To make theoretical predictions, we numerically solved the optical Bloch equations (OBEs) of the density-matrix elements of the atoms, $\rho_{ij}(t,z)$, and Maxwell-Schr\"{o}dinger equations (MSEs) of the forward/backward probe pulses (with the Rabi frequencies of $\Omega_{p+}(t,z)$ and $\Omega_{p-}(t,z)$), where $t$ and $z$ represent the temporal and spatial variables. There are five energy levels in the calculation consisting of two ground states $|1\rangle$ and $|2\rangle$, two excited states $|3\rangle$ and $|4\rangle$, and a Rydberg state $|5\rangle$. The system of the $\Lambda$-type electromagnetically induced transparency (EIT) is formed by $|1\rangle$, $|2\rangle$ and $|3\rangle$. It is driven by the two probe pulses and the forward/backward coupling fields ($\Omega_{c+}(t)$ and $\Omega_{c-}(t)$) to generate the stationary dark-state polariton (DSP)~\cite{SLP}. The coherences of $\rho_{31+}$ and $\rho_{31-}$ correspond to the forward/backward probe transitions, and those of $\rho_{32+}$ and $\rho_{32-}$ to the forward/backward coupling transitions. In the system of the two-photon transition (TPT), the two fields of $\Omega_{a}(t)$ and $\Omega_{b}(t)$ drove the transition of $|2\rangle \rightarrow$ $|4\rangle \rightarrow$ $|5\rangle$ with a large one-photon detuning. The coherences of $\rho_{42}$ and $\rho_{52}$ correspond to the transitions of $\Omega_{a}$ and $\Omega_{b}$, respectively. The TPT excites and deexcites the population between $|2\rangle$ and $|5\rangle$ or, equivalently, converts the coherence of $\rho_{21}$ to that of $\rho_{51}$ and vice versa. 

The OBEs describe how the optical fields influence the density-matrix operator, and they are given by
\begin{widetext}
\begin{eqnarray}
\label{eq:OBEstart}
	\frac{\partial\rho_{21}}{\partial t} = 
		\frac{i}{2}(\Omega^*_{c+}\rho_{31+} +\Omega^*_{c-}\rho_{31-}
		-\Omega_{p+}\rho^{*}_{32+}-\Omega_{p-}\rho^{*}_{32-} +\Omega^*_{a}\rho_{41}) 
		+i\delta_\Lambda \rho_{21} -\gamma_\Lambda\rho_{21},
		\\
	\frac{\partial\rho_{31\pm}}{\partial t} =
		\frac{i}{2}[\Omega_{p\pm}(\rho_{11}-\rho_{33})+\Omega_{c\pm}\rho_{21}]
		+i\Delta_{p\pm}\rho_{31\pm} -\frac{\Gamma_3}{2}\rho_{31\pm},
		\\
	\frac{\partial\rho_{41}}{\partial t} =
		\frac{i}{2}(\Omega_a\rho_{21} +\Omega^*_b\rho_{51} -\Omega_{p+}\rho_{43+}
		-\Omega_{p-}\rho_{43-})
		+i(\Delta_a +\delta_{\Lambda})\rho_{41} -\frac{\Gamma_4}{2}\rho_{41},
		\\
\label{eq:rho51}
	\frac{\partial\rho_{51}}{\partial t} =
		\frac{i}{2}(\Omega_b \rho_{41}-\Omega_{p+}\rho_{53+}-\Omega_{p-}\rho_{53-})
		+i(\delta_R+\Delta_{\rm DDI}+\delta_{\Lambda})\rho_{51}
		-\left( \frac{\Gamma_5}{2} +\gamma_R +\gamma_{\rm DDI}  
		+\gamma_{\Lambda} \right)\rho_{51},
		\\
	\frac{\partial\rho_{32\pm}}{\partial t} =
		\frac{i}{2}[\Omega_{c\pm}(\rho_{22}-\rho_{33})
		+\Omega_{p\pm}\rho^*_{21} -\Omega_a\rho^*_{43\pm}]
		+i\Delta_{c\pm}\rho_{32\pm} -\frac{\Gamma_3}{2}\rho_{32\pm},
		\\
	\frac{\partial\rho_{42}}{\partial t} = 
		\frac{i}{2}[\Omega_a(\rho_{22}-\rho_{44}) +\Omega^*_b\rho_{52}
		-\Omega_{c+}\rho_{43+}-\Omega_{c-}\rho_{43-}]
		+i\Delta_a\rho_{42}-\frac{\Gamma_4}{2}\rho_{42},
		\\
\label{eq:rho52}
	\frac{\partial\rho_{52}}{\partial t} =
		\frac{i}{2}(\Omega_b\rho_{42}-\Omega_a\rho_{54}-\Omega_{c+}\rho_{53+}
		-\Omega_{c-}\rho_{53-}) 
		+i(\delta_R+\Delta_{\rm DDI})\rho_{52}
		-\left( \frac{\Gamma_5}{2}+\gamma_{R}+\gamma_{\rm DDI} \right) \rho_{52},
		\\
	\frac{\partial\rho_{43\pm}}{\partial t}=
		\frac{i}{2}(\Omega_a\rho^*_{32\pm} +\Omega^*_b\rho_{53\pm}
		-\Omega^*_{p\pm}\rho_{41} -\Omega^*_{c\pm}\rho_{42})
		+i(\Delta_a-\Delta_{c\pm})\rho_{43\pm} -\frac{\Gamma_3+\Gamma_4}{2}\rho_{43\pm},
		\\
\label{eq:rho53}
	\frac{\partial\rho_{53\pm}}{\partial t} =
		\frac{i}{2}(\Omega_b\rho_{43\pm} -\Omega^*_{c\pm}\rho_{52}
		-\Omega^*_{p\pm}\rho_{51}) 
		+i(\delta_R+\Delta_{\rm DDI}-\Delta_{c\pm})\rho_{53\pm}
		-\frac{\Gamma_3+\Gamma_5}{2}\rho_{53\pm},
		\\
\label{eq:rho54}
	\frac{\partial\rho_{54}}{\partial t} =
		\frac{i}{2}[\Omega_b(\rho_{44}-\rho_{55}) -\Omega^*_a\rho_{52}]
		+i(\Delta_{\rm DDI} +\Delta_b)\rho_{54} -\frac{\Gamma_4+\Gamma_5}{2}\rho_{54},
		\\
	\frac{\partial\rho_{11}}{\partial t} =
		\frac{i}{2}(\Omega^*_{p+}\rho_{31+}-\Omega_{p+}\rho^*_{31+}
		+\Omega^*_{p-}\rho_{31-}-\Omega_{p-}\rho^*_{31-}) +\frac{\Gamma_3}{2}\rho_{33},
		\\
	\frac{\partial\rho_{22}}{\partial t} =
		\frac{i}{2}(\Omega^*_{c+}\rho_{32+} -\Omega_{c+}\rho^*_{32+}
		+\Omega^*_{c-}\rho_{32-} -\Omega_{c-}\rho^*_{32-} +\Omega_a^*\rho_{42}
		-\Omega_a\rho^*_{42})+\frac{\Gamma}{2}\rho_{33} +\Gamma_4\rho_{44},
		\\
	\frac{\partial\rho_{44}}{\partial t} =
		\frac{i}{2}(\Omega_a\rho^*_{42} -\Omega_a^*\rho_{42} +\Omega^*_b\rho_{54}
		-\Omega_b\rho^*_{54})-\Gamma_4\rho_{44},
		\\
\label{eq:OBEend}
	\frac{\partial\rho_{55}}{\partial t} =
		\frac{i}{2}(\Omega_b\rho^*_{54}-\Omega^*_b\rho_{54}) -\Gamma_5\rho_{55},
\end{eqnarray}
\end{widetext}
where $\Delta_{c\pm}$ and $\Delta_{p\pm}$ are the one-photon detunings of $\Omega_{c\pm}$ and $\Omega_{p\pm}$, respectively, $\delta_\Lambda$  ($= \Delta_{p+} -\Delta_{c+} = \Delta_{p-} -\Delta_{c-}$) is the two-photon detuning of the $\Lambda$-type EIT transitions, $\gamma_{\Lambda}$ is the ground-state decoherence rate, $\Delta_a$ and $\Delta_b$ are the one-photon detunings of $\Omega_a$ and $\Omega_b$, respectively, $\delta_R$ ($= \Delta_a + \Delta_b$) is the two-photon detuning of the TPT, $\gamma_R$ is the Rydberg-state decoherence rate, $\Delta_{\rm DDI}$ and $\gamma_{\rm DDI}$ are the DDI-induced frequency shift and decoherence rate, and $\Gamma_3$ ($= \Gamma$ or 2$\pi$$\times$6 MHz), $\Gamma_4$ ($= \Gamma$), and $\Gamma_5$ (= 1.3$\times$$10^{-3}$$\Gamma$) represent the spontaneous decay rates of $|3\rangle$, $|4\rangle$, and $|5\rangle$, respectively. 

Since $\gamma_\Lambda$, $\gamma_R$, $\gamma_{\rm DDI}$ $\ll$ $\Gamma$ in the experiment, these three decoherence rates do not appear in the time-derivative equations of $\rho_{53\pm}$ and $\rho_{54}$ in the calculation. Furthermore, $\Gamma_5 \ll \Gamma$, $\Gamma_5 \ll \gamma_R$, and $\Gamma_5^{-1} \gg$ the TPT duration in the experiment. Thus, we set $\Gamma_5 = 0$ in the calculation, which did not change the outcome. The forward probe and coupling fields had $\Delta_{c+} = \Delta_{p+} \equiv \Delta_+ \approx 0$, and the backward probe and coupling fields had $\Delta_{c-} = \Delta_{p-} \equiv \Delta_- = -1.0$$\Gamma$. Such an arrangement of the one-photon detunings suppressed possible high-order coherences, which could be established by the counter-propagating fields of $\Omega_{p+}$ and $\Omega_{c-}$ ($\Omega_{p-}$ and $\Omega_{c+}$)~\cite{SLP}. Consequently, we do not need to consider the high-order coherences in the calculation.

In the present work, the experimental condition satisfied the weak-interaction regime, i.e., $(r_B/r_a)^3 \ll 1$ where $r_B$ is the blockade radius and $r_a$ is the half mean distance between Rydberg atoms. Hence, $\Delta_{\rm DDI}$ and $\gamma_{\rm DDI}$ in the OBEs are given by~\cite{OurOEMFT, OurCommunPhys2021, OurPRR2022} 
\begin{eqnarray}
\label{eq:dDDI}
	\Delta_{\rm DDI}= A \rho_{55}, \\
\label{eq:gDDI}
	\gamma_{\rm DDI}= A \rho_{55},
\end{eqnarray}
where $A$ represents the frequency shift (also the decoherence rate) per unit population of $|5\rangle$, and its value depends on the atomic density in the system and the DDI strength. The appearances of $\Delta_{\rm DDI}$ and $\gamma_{\rm DDI}$ make the OBEs nonlinear. The OBEs involving the DDI effect are very similar to those in our previous works of Refs.~\cite{OurCommunPhys2021, OurPRR2022}, which were derived based on the mean field model studied in Ref.~\cite{OurOEMFT}. In those works, the theoretical predictions are in agreement with the experimental data.

The MSEs describe how the forward and backward probe fields are affected by the atomic coherences, and they are given by
\begin{eqnarray}
\label{eq:MSEforward}
	\frac{1}{c}\frac{\partial}{\partial t} \Omega_{p+} +\frac{\partial}{\partial z}\Omega_{p+}
 		= i\frac{\alpha \Gamma}{2L} \rho_{31+},
		\\
\label{eq:MSEbackward}
	\frac{1}{c}\frac{\partial}{\partial t} \Omega_{p-} -\frac{\partial}{\partial z}\Omega_{p-}
		+i\Delta_k\Omega_{p-} 
		=  i\frac{\alpha \Gamma}{2L} \rho_{31-},
\end{eqnarray}
where $\alpha$ and $L$ are the optical depth and length of the atom cloud, and $\Delta_k$ $\equiv (\vec{k}_{p-}-\vec{k}_{c-}-\vec{k}_{p+}+\vec{k}_{c+})$$\cdot$$\hat{z}$ and $\vec{k}_{p\pm}$ and $\vec{k}_{c\pm}$ are the wave vectors of $\Omega_{p\pm}$ and $\Omega_{c\pm}$, respectively. The value of $L\Delta_k$ is the degree of phase mismatch in the system. Although $\Delta_k$ only appears in the MSE of $\Omega_{p-}$, it reduces the energies of $\Omega_{p+}$ and $\Omega_{p-}$ equally due to the coupling between the forward and backward fields, i.e., the four-wave mixing process~\cite{Ldk2004, SLP2012, FBS2021}. 

\section{Illustration of Figures 2(a)and 2(b) of the main text}
\label{sec:attenprocedure}

We utilize~\MainFigTwo~to illustrate the procedure and method for forming the stationary DSPs and driving the two-photon transition (TPT) from the ground state to the Rydberg state. In the figures, the black and red circles are the experimental data of the input and output probe signals as functions of time, the red lines are the theoretical predictions, the black line is the Gaussian function used as the input probe pulse in the theoretical calculation, and the blue, green, magenta, and cyan lines represent the timing sequences of the forward coupling (the Rabi frequency denoted as $\Omega_{c+}$), backward coupling ($\Omega_{c-}$), and two TPT fields ($\Omega_a$ and $\Omega_b$), respectively. The theoretical predictions were calculated numerically with the equations presented in Sec.~\ref{sec:Theory}, and they are consistent with the experimental data well.

In~\MainFigTwoA, we formed the stationary DSP and did not drive the TPT. The duration before $t \approx$ 2 $\mu$s is the Input period specified on the figure.
The probe pulse with the peak Rabi frequency $\Omega_{p0}$ was sent into the atom cloud in the forward direction under the presence of only $\Omega_{c+}$ ($= 0.54$$\Gamma$). As the majority of the probe pulse had entered the atom cloud and propagated slowly, we quickly switched off $\Omega_{c+}$ as indicated by the falling edge of the blue line at $t \approx$ 2~$\mu$s. At this moment, the part of the pulse centering around its peak inside the cloud was stored. 
This is the end of the Input period.
The black circles after $t \approx$ 2 $\mu$s and the red circles before $t \approx$ 2 $\mu$s were the parts of the pulse outside the cloud, and they were not stored.

After the Input period shown on~\MainFigTwoA, there was a fixed storage time. Then, the period of Stationary DSP started.
We simultaneously switched on $\Omega_{c+}$ and $\Omega_{c-}$ (both are equal to 0.44$\Gamma$), as indicated by the green and blue pulses, to form the stationary DSP for a duration of about 0.7~$\mu$s. Within the stationary-DSP duration, the leakage of the probe pulse in the forward direction is shown by the red circles underneath the blue square pulse. There was also the leakage in the backward direction, but we did not measure it. After keeping $\Omega_{c+}$ and $\Omega_{c-}$ on for about 0.7~$\mu$s, we switched them off. 
This is the end of the Stationary DSP period.
The remaining probe energy was stored in the atoms for another fixed time. Next, the Output period started.
We switched on $\Omega_{c+}$ (= 0.54$\Gamma$) again and retrieved the probe pulse. The red circles after the rising edge of the blue line at $t \approx$ 6.5~$\mu$s represent the retrieved probe pulse. Other details of forming the stationary DSP can be found in Ref.~\cite{SLP}.

In~\MainFigTwoB, we formed the stationary DSP and also drove the TPT simultaneously. Since the Input period here was completely the same as that of~\MainFigTwoA, it is not drawn again.
The TPT fields $\Omega_a$ (= 2.0$\Gamma$) and $\Omega_b$ (= 1.6$\Gamma$) were applied to induce the DDI during the Stationary DSP period. The field $\Omega_a$ is indicated by the magenta pulse. 
Without the presence of $\Omega_a$, the field of $\Omega_b$ does not affect the effects of EIT, slow light, and storage of light at all. Hence, for simplicity it was constantly present.

\MainFigTwoBB~shows the representative data of the stationary DSP dressed by the DDI, and the magenta square pulse corresponds to the TPT duration. The TPT fields produced the Rabi oscillation between $|2\rangle$ and $|5\rangle$. Consequently, in the figure the red circles of the probe leakage within the duration exhibit the oscillation. The node (anti-node) of the oscillation corresponds to all the population in $|5\rangle$ (in $|2\rangle$). 
We set the TPT frequency to the two-photon resonance by 
minimizing the anti-node.
Due to the DDI-induced attenuation, the retrieved probe energy in~\MainFigTwoB~is obviously less than that in~\MainFigTwoA.

\section{Determination of the Rabi Frequencies }
\label{sec:RabiFreq}

In this section, we illustrate the methods that determine the Rabi frequencies of the forward and backward coupling fields $\Omega_{c+}$ and $\Omega_{c-}$, and that of the two-photon-transition (TPT) field $\Omega_{a}$. We also checked whether the Rabi frequency of the TPT field $\Omega_b$ determined by the data in \MainFigTwoB~is reasonable.  

The Autler-Townes splitting (ATS) in the EIT transmission spectrum was used to determine the Rabi frequencies of $\Omega_{c+}$ and $\Omega_{c-}$. Figure~\ref{fig:ATS} shows the representative EIT spectra. We followed the determination method described in Ref.~\cite{OurPRA2019}. The frequency difference between two transmission minima, $\delta_+$ and $\delta_-$, gave the Rabi frequency. As mentioned in the reference, the optical depth (OD) of the atom cloud was reduced to a value between 1 and 2 in the measurement of the spectrum. At a low OD, the minima can be clearly identified, and any asymmetry in the spectrum due to a nonzero one-photon detuning of the coupling field can be noticed. In the EIT system, 
\begin{equation}
	\delta_+ -\delta_- \approx \Omega + \Delta^2 /2\Omega, 
\end{equation}
where $\Omega$ is the coupling Rabi frequency and $\Delta$ is the one-photon detuning. As long as $\Delta^2 \ll \Omega^2$, the angular-frequency separation between $\delta_+$ and $\delta_-$ is nearly the same as $\Omega$.

\FigSOne

In the $\Lambda$-type EIT system, we employed the forward probe field ($\Omega_{p+}$) to measure the values of the forward and backward coupling Rabi frequencies, $\Omega_{c+}$ and $\Omega_{c-}$, as shown in Figs.~\ref{fig:ATS}(a) and \ref{fig:ATS}(b), respectively. The sweeping of the probe frequency deviated the probe propagation direction very little. Since the propagation direction of $\Omega_{c-}$ was opposite to that of $\Omega_{p+}$, the Doppler effect made the decoherence rate in Fig.~\ref{fig:ATS}(b) much larger than that in Fig.~\ref{fig:ATS}(a). Consequently, the EIT peak transmission in Fig.~\ref{fig:ATS}(b) is significantly reduced, and that in Fig.~\ref{fig:ATS}(a) is nearly 100\%. Nevertheless, the frequency separation between the two transmission minima is not affected by the decoherence rate.

\FigSTwo

To determine the value of $\Omega_a$ of the TPT field, we utilized the all-optical switching (AOS) effect in the four-level $N$-type system~\cite{AOS, XPM}, which consists of states $|1\rangle$, $|2\rangle$, $|3\rangle$, and $|4\rangle$ in this work. In the AOS, the forward probe ($\Omega_{p+}$) and coupling ($\Omega_{c+}$) fields form the $\Lambda$-type EIT configuration, and the presence of the field of $\Omega_a$ causes an additional attenuation of the output probe field. As shown in the inset of Fig.~\ref{fig:AOS}, we first stored the probe pulse in the atoms, then applied the pulse of $\Omega_a$ during the storage time, and finally retrieved the probe pulse. The one-photon detuning, $\Delta_a$, of the $\Omega_a$ pulse was kept to $+5.0$$\Gamma$ in the measurement. Except that the fields of $\Omega_{c-}$ and $\Omega_b$ were absent, all the other experimental arrangements were the same as those in the measurement with the TPT. In Fig.~\ref{fig:AOS}, the circles and black line are the experimental data and best fit of the retrieved probe energy transmission, $T_{\rm AOS}$, as a function of the $\Omega_a$ pulse duration, $\tau_a$. The fitting function is given by
\begin{equation}
\label{eq:AOS}
	T_{\rm AOS} = T_0 \exp\left( -\frac{\Omega_a^2 \Gamma}{\Gamma^2+4\Delta_a^2} 
		\tau_a \right),
\end{equation}
where $T_0$ is the transmission without the presence of $\Omega_a$. The best fit determined the value of $\Omega_a$ used in the TPT.

We measured the ATS induced by the TPT field $\Omega_b$ in a Rydberg-EIT system~\cite{OurPRA2019}. In the system, the coupling field $\Omega_b$ and a weak probe field drove the transition of $|5P_{3/2},F=3,m_F=3\rangle$ $\rightarrow$ $|32D_{5/2},m_J=5/2\rangle$ and that of $|5S_{1/2},F=2, m_F=2\rangle$ $\rightarrow$ $|5P_{3/2},F=3,m_F$=$3\rangle$, respectively, and the two fields counter-propagated. The input probe Rabi frequency was set to about 0.05$\Gamma$ such that the DDI effect was negligible. Figure~\ref{fig:ATS}(c) shows the Rydberg-EIT spectrum taken at a nearly zero one-photon detuning. The ATS in the spectrum is 2.0$\Gamma$. However, the transition of $\Omega_b$ in Fig.~\ref{fig:ATS}(c) was not the same as that of $\Omega_b$ in the TPT, which was from $|5P_{3/2},F=3,m_F=2\rangle$ to $|32D_{5/2},m_J=3/2~\&~5/2\rangle$. According to the ATS and the Clebsch-Gordan coefficients, the Rabi frequency of the transition to $|32D_{5/2},m_J=5/2\rangle$ and that to $|32D_{5/2},m_J=3/2\rangle$ should be 1.4$\Gamma$ and 1.1$\Gamma$, respectively. Their root-mean-square value is 1.8$\Gamma$. It is reasonable that the TPT field $\Omega_b$ had the Rabi frequency of 1.6$\Gamma$ as determined by \MainFigTwoB.

\section{Degree of Phase Mismatch During the Stationary Dark-State Polariton}
\label{sec:SDSP}

The stationary dark-state polariton (DSP) was formed by the balance between the two four-wave mixing (FWM) processes of $\Omega_{p+} - \Omega_{c+} + \Omega_{c-} \rightarrow \Omega_{p-}$ and $\Omega_{p-} - \Omega_{c-} + \Omega_{c+} \rightarrow \Omega_{p+}$~\cite{SLP2012}, where $\Omega_{p\pm}$ and $\Omega_{c\pm}$ denote the forward/backward probe and coupling fields, respectively. The former process consists of absorbing a forward probe photon, emitting a forward coupling photon, and absorbing a backward coupling photon to finally generate a backward probe photon. The latter process is just the reversed former process to generate a forward probe photon from a backward probe photon. In the FWM, the phase mismatch causes an additional attenuation of the stationary DSP, and its measure, i.e., the degree of the phase mismatch, is given by $L\Delta_k$, where $L$ is the medium length and $\Delta_k$ represents the sum of the wave vectors of the four fields projected on the propagation direction. Please see Eqs.~(\ref{eq:MSEforward}) and (\ref{eq:MSEbackward}) for the role of $L\Delta_k$ in the Maxwell-Schr\"{o}dinger equations. The definition and details of $\Delta_k$ are given right after the equations.

To determine the value of $L\Delta_k$ in the stationary-DSP experiment, we measured the retrieved probe energy as a function of the stationary-DSP duration. The experimental condition or parameters of the measurement was the same as that of \MainFigTwoA. As shown in the inset of Fig.~\ref{fig:StationaryDSP}, we first stored the probe pulse in the atoms, then applied the pulse of the forward and backward coupling fields to generate the stationary DSP for a given duration, $\tau_{\rm DSP}$, and finally retrieved the probe pulse after the execution of $\tau_{\rm DSP}$. The time length from storing to retrieving the probe pulse was fixed to about 10~$\mu$s to accommodate the variation of the stationary-DSP duration. The main plot of Fig.~\ref{fig:StationaryDSP} shows the retrieved probe energy transmission versus $\tau_{\rm DSP}$ at three different values of the peak Rabi frequency, $\Omega_{p0}$, of the input probe pulse. In the figure, the circles are the experimental data. The retrieved probe energy transmissions of $\Omega_{p0} =$ 0.07$\Gamma$ and 0.24$\Gamma$ are about the same, but is higher than that of $\Omega_{p0} =$ 0.40$\Gamma$. As $\Omega_{p0} \leq$ 0.24$\Gamma$, the probe pulse is weak enough to be treated as the perturbation, and the result does not depend on $\Omega_{p0}$. On the other hand, as $\Omega_{p0}$ becomes larger, the probe pulse is no longer the perturbation and its intensity influences the outcome. 

\FigSThree

In Fig.~\ref{fig:StationaryDSP}, the colored areas represent the theoretical predictions. We used the experimentally-determined values of the optical density ($\alpha$), the forward and backward coupling Rabi frequencies ($\Omega_{c+}$ and $\Omega_{c-}$), and the ground-state decoherence rate ($\gamma_{\Lambda}$) in the theoretical calculation. The determination methods of $\alpha$ and $\gamma_{\Lambda}$ can be found in the last paragraph of \SecExpSetup, and that of $\Omega_{c+}$ and $\Omega_{c-}$ in the second and third paragraphs of \SecRabiFreq. Due to the fluctuations in $\alpha$ and $\gamma_{\Lambda}$, the prediction of each $\Omega_{p0}$ is displayed as an area. A single value of $L\Delta_k$ of 0.9 rad was used in all the predictions such that the consistency between the data and predictions is satisfactory as shown in Fig.~\ref{fig:StationaryDSP}. This nonzero $L\Delta_k$ can be caused by a little misalignment among the fields in the FWM processes.

\section{Estimation of the Coefficient \textit{A} of Dipole-Dipole Interaction}

We estimate the strength of dipole-dipole interaction (DDI) or the coefficient $A$ with a simplified model in this section, where $A$ is defined as the DDI-induced decoherence rate, $\gamma_{\rm DDI}$, and frequency shift, $\Delta_{\rm DDI}$, per unit Rydberg-state population, $\rho_{55}$, as shown by Eqs.~(\ref{eq:dDDI}) and (\ref{eq:gDDI}). The coefficient $A$ was used in the equations of the coherences $\rho_{51}$, $\rho_{52}$, $\rho_{53}$, and $\rho_{54}$ relating to the Rydberg state $|5\rangle$, i.e., Eqs.~(\ref{eq:rho51}), (\ref{eq:rho52}), (\ref{eq:rho53}), and (\ref{eq:rho54}), respectively. In the experiment, we utilized a series of measurements of the output probe energy transmission against the input probe intensities or $\Omega_{p0}^2$ to determine $A =$ 0.60$\Gamma$ as shown by \MainFigTwo. 

The formation of stationary dark-state polaritons (DSP) is not relevant to the DDI strength. Consequently, state $|3\rangle$ and the probe and coupling fields are not present in the model here. To focus on the DDI effect, we only consider states $|1\rangle$, $|2\rangle$, and $|5\rangle$. An effective field with the Rabi frequency of $\Omega_{\rm eff}$ drives the one-photon transition (OPT) between $|2\rangle$ and $|5\rangle$. Due to a large one-photon detuning, $\Delta_a$, in the experiment, the intermediate state $|4\rangle$ of the two-photon transition (TPT) can be eliminated. Thus, the TPT of $|2\rangle \leftrightarrow |4\rangle \leftrightarrow |5\rangle$ with the driving fields of $\Omega_a$ and $\Omega_b$ in the experiment is equivalent to the OPT of $|2\rangle \leftrightarrow |5\rangle$ here, and 
\begin{equation}
	\Omega_{\rm eff} \approx \frac{\Omega_a\Omega_b}{2\Delta_a}.
\end{equation}
Considering $\Omega_a =$ 2.0$\Gamma$, $\Omega_b =$ 1.6$\Gamma$, and $\Delta_a =$ 5.0$\Gamma$ of the TPT in the experiment, $\Omega_{\rm eff} =$ 0.32$\Gamma$ in the OPT. 

The OPT drives the Rabi oscillation between $|2\rangle$ and $|5\rangle$, and the equations of motion are given by
\begin{eqnarray}
\label{eq:OBEopt1}
	\frac{\partial\rho_{55}}{\partial t} =
		\frac{i}{2}(\Omega_{\rm eff}\rho^*_{52}-\Omega^*_{\rm eff}\rho_{52}),
		\\
	\frac{\partial\rho_{52}}{\partial t} =
		\frac{i}{2}\Omega_{\rm eff}(\rho_{22}-\rho_{55}) 
		+iA\rho_{55}\rho_{52} -A\rho_{55}\rho_{52},
		\\
\label{eq:OBEopt3}
	\rho_0 = \rho_{22} + \rho_{55},
\end{eqnarray}
where $A\rho_{55}$ is substituted for $\gamma_{\rm DDI}$ and $\Delta_{\rm DDI}$ and $\rho_0$ is the initial total population in the system. In the above equations, we neglect the spontaneous decay rate of $|5\rangle$ (i.e., $\Gamma_5$) and all the other decoherence rates (i.e., $\gamma_\Lambda$ and $\gamma_R$), which appear in Eqs.~(\ref{eq:OBEstart})-(\ref{eq:OBEend}).  

The OPT also converts the coherences $\rho_{21}$ and $\rho_{51}$ back and forth, and the equations of the conversion are given by
\begin{eqnarray}
	\frac{\partial\rho_{21}}{\partial t} = \frac{i}{2}\Omega_{\rm eff}\rho_{51},
		\\
	\frac{\partial\rho_{51}}{\partial t} = \frac{i}{2} \Omega_{\rm eff} \rho_{21} 
		+i \omega\rho_{51} -\gamma\rho_{51},
\end{eqnarray}
where $\omega$ and $\gamma$ correspond to the detuning and decoherence rate of the OPT. Since the output probe energy transmission is proportional to $|\rho_{21}|^2$, we are interested in the solution of $\rho_{21}$ given by
\begin{widetext}
\begin{eqnarray}
	{\rm Re}[\rho_{21}(\omega,\gamma)] &\approx& \sqrt{\rho_0}
		e^{-\gamma t/2} \left[ 
		\cos\left(\frac{\omega t}{2}\right) \cos\left(\frac{\Omega' t}{2}\right)
		+ \frac{\omega}{\Omega'} 
		\sin\left(\frac{\omega t}{2}\right) \sin\left(\frac{\Omega' t}{2}\right)
		+ \frac{\gamma}{\Omega'} 
		\cos\left(\frac{\omega t}{2}\right) \sin\left(\frac{\Omega' t}{2}\right)
		 \right],
		\\
	{\rm Im}[\rho_{21}(\omega,\gamma)] &\approx& \sqrt{\rho_0}
		e^{-\gamma t/2} \left[ 
		\sin\left(\frac{\omega t}{2}\right) \cos\left(\frac{\Omega' t}{2}\right)
		- \frac{\omega}{\Omega'} 
		\cos\left(\frac{\omega t}{2}\right) \sin\left(\frac{\Omega' t}{2}\right)
		+ \frac{\gamma}{\Omega'} 
		\sin\left(\frac{\omega t}{2}\right) \sin\left(\frac{\Omega' t}{2}\right)
		 \right],
		\\
\label{eq:rho21sq}
    |\rho_{21}(\omega,\gamma)|^2 &\approx& \rho_0 
        e^{-\gamma t} \left\{ 
        \frac{1}{2} [1+\cos(\Omega' t)]
        + \frac{\gamma}{\Omega'} \sin(\Omega' t)
        + \frac{\gamma^2}{2 \Omega'^2} [1-\cos(\Omega' t)]
        + \frac{\omega^2}{2 \Omega'^2} [1-\cos(\Omega' t)] \right\},
		\\
	\Omega' &\equiv& \sqrt{\Omega_{\rm eff}^2+\omega^2}.
\end{eqnarray}
\end{widetext}

To estimate the value of $A$, we utilized the mean-field approach based on the probability function of nearest-neighbor distribution (NND)~\cite{OurOEMFT, NNDistribution}. Considering Rydberg atoms are randomly distributed, the probability of finding the DDI-induced frequency shift within $\omega$ and $\omega +d\omega$ is given by
\begin{eqnarray}
	P(\omega) \!\!\!&&= \frac{1}{\omega_a}
		\frac{\left[ 1+\sqrt{1+4(\omega/\omega_a)} \right]^2}
		{4 (\omega/\omega_a)^2 \sqrt{1+4(\omega/\omega_a)}}
		\nonumber \\
 	&& \times
		\exp{\left[ -\frac{1+\sqrt{1+4(\omega/\omega_a)}}{2(\omega/\omega_a)} \right]},	
\label{eq:Pomega}
\end{eqnarray}
where
\begin{equation}
		\omega_a \equiv |C_6| /r_a^6,
\label{eq:Define_omega_a}
\end{equation}
$C_6$ is the van der Waals coefficient, $r_a$ is the half-mean distance between the Rydberg atoms given by
\begin{equation}
	(4\pi/3) r_{a}^3 \equiv (n_{\rm atom}\rho_{55})^{-1},
\label{eq:Define_r_a}
\end{equation}
and $n_{\rm atom}$ is the atomic density. The derivation of $P(\omega)$ using the NND can be found in Ref.~\cite{OurOEMFT}. We performed the following integrals to simulate the DDI effect:
\begin{eqnarray}
\label{eq:bar_rho21r}
	\bar{\rho}_{21,r}(t) &=& \int_0^\infty d\omega P(\omega)~
		{\rm Re}[\rho_{21}(\omega, \gamma=0)], 
		\\ 
\label{eq:bar_rho21i}
	\bar{\rho}_{21,i}(t) &=& \int_0^\infty d\omega P(\omega)~
		{\rm Im}[\rho_{21}(\omega, \gamma=0)],
		\\
\label{eq:bar_rho21sq}
	|\bar{\rho}_{21}|^2 &\equiv& [\bar{\rho}_{21,r}(t)]^2 + [\bar{\rho}_{21,i}(t)]^2.
\end{eqnarray}
Because of the distribution function $P(\omega)$, $|\bar{\rho}_{21}|^2$ is a function of $\rho_{55}$ as indicated by Eqs.~(\ref{eq:Pomega})-(\ref{eq:Define_r_a}).

\FigSFour

The process of the estimation of $A$ is illustrated as the followings: (i) We set $A = 0$ and numerically solved Eqs.~(\ref{eq:OBEopt1})-(\ref{eq:OBEopt3}) to obtain $\rho_{55}(t)$ as shown by the black circles in Fig.~\ref{fig:EstimateA}(a). (ii) With $\rho_{55}(t)$, we used Eqs.~(\ref{eq:bar_rho21r})-(\ref{eq:bar_rho21sq}) to evaluate the temporal data points of $|\bar{\rho}_{21}|^2$ represented by the black circles in Fig.~\ref{fig:EstimateA}(b). (iii) The data points in Fig.~\ref{fig:EstimateA}(b) are fitted with the function of $|\rho_{21}(\omega =A\rho_{55}, \gamma = A\rho_{55})|^2$ shown by Eq.~(\ref{eq:rho21sq}), where $A$ is the only fitting parameter. The fitting means to relate the ensemble average of the microscopic DDI effect to the macroscopic quantity $A$, i.e., 
\begin{equation}
	 |\bar{\rho}_{21}|^2 = |\rho_{21}(\omega =A\rho_{55}, \gamma = A\rho_{55})|^2.
\end{equation}
The black line in Fig.~\ref{fig:EstimateA}(b) is the best fit, which determined the value of $A$ in the first iteration, denoted as $A^{(1)}$. (iv) In the second iteration, we started with the value of $A^{(1)}$ in Eqs.~(\ref{eq:OBEopt1})-(\ref{eq:OBEopt3}) and solved the equations to obtain $\rho_{55}^{(2)}(t)$ as shown by the red circles in Fig.~\ref{fig:EstimateA}(a). Steps (ii) and (iii) were repeated, and the red circles and line in Fig.~\ref{fig:EstimateA}(b) are the results, giving $A^{(2)}$. (v) The process of (i) $\rightarrow$ (ii) $\rightarrow$ (iii) was performed several times until the results converged as demonstrated by the blue squares/line and green diamonds/line in Figs.~\ref{fig:EstimateA}(a) and \ref{fig:EstimateA}(b).

In the estimation, we set $C_6 =$ --2$\pi$$\times$130~MHz$\cdot$$\mu$m$^6$, $n_{\rm atom} = 0.05$ $\mu$m$^{-3}$, and $\Omega_{\rm eff} =$ 0.32$\Gamma$ according to the experimental condition. Considering the population of $|2\rangle$ achievable in this work, the value of $A$ estimated by the method in this section depends on $\rho_0$ very little. As the estimation gives $A$ of 0.54$\Gamma$, the experimentally determined $A$ of 0.60$\Gamma$ sounds reasonable.

\section{The Input Probe Pulse with a Large Rabi Frequency}

We employed the input probe pulses with five different peak Rabi frequencies, $\Omega_{p0}$, of 0.07$\Gamma$, 0.2$\Gamma$, 0.24$\Gamma$, 0.3$\Gamma$, 0.4$\Gamma$ in the experiment. As $\Omega_{p0} \leq$ 0.24$\Gamma$, the experimental data of slow light were well consistent with the theoretical predictions. As $\Omega_{p0} \geq$ 0.3$\Gamma$, the experimental data of slow light deviated from the theoretical predictions. A representative example of the discrepancy taken at $\Omega_{p0}$ of 0.4$\Gamma$ is shown in Fig.~\ref{fig:LargeProbe}(a). In the theoretical calculations, we used the parameters of the coupling Rabi frequency ($\Omega_{c+}$), optical depth or OD ($\alpha$), and ground-state decoherence rate ($\gamma_\Lambda$), which were determined experimentally by the methods described in \SecRabiFreq~and in the last paragraph of \SecExpSetup. The determinations were carried out at the experimental condition that the typical electromagnetically induced transparency (EIT) criterion, i.e., $\Omega_{p0}^2 \ll \Omega_{c+}^2$ and $2\gamma\Gamma \ll \Omega_{c+}^2$, is satisfied. 

\FigSFive

The comparison between the experimental data and theoretical predictions in Fig.~\ref{fig:LargeProbe}(a) leads to the following observations: (i) The experimental output pulse peak is smaller than the theoretical one, (ii) the experimental output pulse width is broader than the theoretical one, and (iii) the experimental propagation delay time is approximately the same as the theoretical one. These observations can explain the results of the output probe energies without and with the two-photon transition (TPT).

Figure~\ref{fig:LargeProbe}(b) shows the input and output probe pulses without the presence of the TPT. The timing sequences of the forward ($\Omega_{c+}$) and backward ($\Omega_{c-}$) coupling fields are nearly the same as those shown in \MainFigTwoA. As we switched off $\Omega_{c+}$ at $t \approx$ 2.0 $\mu$s, the stored probe energy of the experimental data was less than that of the theoretical prediction due to the observation (i). This first factor could make the experimental probe energy less than the predicted one at the output stage, i.e., $t \gtrsim $ 6.6 $\mu$s. Nevertheless, the stored probe pulse shape of the experimental data had a smoother spatial variation than that of the theoretical prediction due to the observation (ii). A smoother pulse shape indicates a narrower frequency bandwidth, which makes the energy loss less during the formation of stationary dark-state polaritons (DSPs)~\cite{SLP, SLPCJP}, i.e., the duration of the green plus and blue pulse in Fig.~\ref{fig:LargeProbe}(b). A narrower frequency bandwidth also makes the energy loss less during the pulse propagation at the output stage~\cite{OurPRL2013}. This second factor could make the experimental output probe energy more than the predicted. Finally, the medium of EIT plus stationary DSP can be viewed as a frequency plus spatial bandpass filter. This third factor can make the output probe pulse shape is approximately independent of the input one. Putting the three factors together, we can expect that the experimental and theoretical output probe pulses are consistent as demonstrated by Figs.~\ref{fig:LargeProbe}(b) and \ref{fig:LargeProbe}(c), and also by the rightmost blue- and red-circle data points and their corresponding predictions in \MainFigTwoC.

When the probe Rabi frequency is comparable to the coupling Rabi frequency as well as the spontaneous decay rate in the EIT system, the EIT effect is no longer independent of the probe Rabi frequency. We think a three-dimensional calculation of the optical Bloch equations and the Maxwell-Schr\"{o}dinger equation is necessary to make a precise theoretical prediction of the slow light. Nevertheless, the three-dimensional calculation is beyond the scope of this work. The discrepancy between the experimental data and theoretical prediction at a large input probe Rabi frequency can be explained, and it does not affect the conclusion of this work at all.

\section{The DDI-Induced Phase Shift under the Rabi Oscillation}

When we applied the two-photon transition (TPT) in the experiment, it not only produced the Rabi oscillation between the ground state $|2\rangle$ and the Rydberg state $|5\rangle$ but also induced the dipole-dipole interaction (DDI). The Rabi oscillation alone resulted in a phase shift of the ground-state coherence $\rho_{21}$, and the DDI added an extra phase shift to $\rho_{21}$. Note that the phase shift of $\rho_{21}$ was transferred to that of the output probe pulse after the retrieval~\cite{XPM}. 

\FigSSix 

Figure~\ref{fig:PhaseShift}(a) shows the theoretical predictions, elucidating the phase shift of $\rho_{21}$ induced by the Rabi oscillation and also that induced by the DDI. To enhance the DDI-induced phase shift, the decay rates $\gamma_\Lambda$, $\gamma_R$, $\gamma_{\rm DDI}$, and $\Gamma_4$ as well as the degree of phase mismatch $L\Delta_k$ were set to zero in the calculation of the predictions. We varied the value of $A$ to switch off or on the DDI-induced phase shift,  i.e., DDI-induced frequency shift or $\Delta_{\rm DDI} = A \rho_{55}$. In Fig.~\ref{fig:PhaseShift}(a), the black solid line represents the result of $A = 0$, i.e., the DDI is switched off, and it reveals the oscillation of the phase shift, which is the consequence of the Rabi oscillation. The black dashed is the baseline of the oscillation. At $A =$ 0.60$\Gamma$, the red solid line represents the result of the phase shift produced by the DDI together with the Rabi oscillation. The red dashed is the baseline of the oscillation. Thus, the phase difference between the red and black dashed lines is solely caused by the DDI. As the interaction time gets longer, the DDI-induced phase shift becomes larger as expected.

We next consider a more realistic case by restoring all the decay terms in the calculation. The values of the decay rates $\gamma_\Lambda$, $\gamma_R$, $\Gamma_4$, and $L\Delta_k$ were set to those specified in the caption of \MainFigTwo, and the DDI-induced attenuation was brought back, i.e., $\gamma_{\rm DDI} = A \rho_{55}$. In Fig.~\ref{fig:PhaseShift}(b), the black and red solid lines are the results without and with the DDI of $A =$ 0.60$\Gamma$. The difference between the black and red dashed lines corresponds to the DDI-induced phase shift. Due to the decay terms, the amount of the DDI-induced phase shift is reduced. 

In the simulations for the results shown in Figs.~\ref{fig:PhaseShift}(a) and \ref{fig:PhaseShift}(b), we made the two TPT fields have the same Rabi frequency, i.e., $\Omega_a = \Omega_b =$ 2.0$\Gamma$, and set the backward coupling field frequency to the resonance of its transition, i.e., $\Delta_- = 0$. The remaining calculation parameters are the same as the experimental parameters shown in the caption of \MainFigTwo. In the experiment, $\Omega_a$ (= 2.0$\Gamma$)  $\neq \Omega_b$ (= 1.6$\Gamma$) and $\Delta_- = -1$$\Gamma$. Both $\Omega_a \neq \Omega_b$ and $\Delta_- = -1$$\Gamma$ can induce AC Stark shifts, which produces a phase shift to $\rho_{21}$. Nevertheless, the phase shifts produced by the causes of the Rabi oscillation, $\Omega_a \neq \Omega_b$, and $\Delta_- = -1$$\Gamma$ depend on the input probe Rabi frequency a little. In \MainFigThreeB, the data points are the results that the measured phase shifts at different values of the input probe Rabi frequency, $\Omega_{p0}$, were subtracted by that at the smallest value of $\Omega_{p0}$. Therefore, the phase shifts revealed by these data points are mainly contributed from the DDI. 

\section{Estimations of the BEC transition temperature, stationary-DSP temperature, and elastic collision rate}

In Ref.~\cite{Fleischhauer2008}, Fleischhauer {\it et al}. proposed to realize the Bose-Einstein condensation (BEC) with stationary dark-state polaritons (DSP). We employ the formulas given by Ref.~\cite{Fleischhauer2008} to estimate the BEC transition temperature and the stationary-DSP temperature, and use the hard-sphere model to evaluate the elastic collision rate in this section. The estimations are based on the experimental condition in the present work. The BEC transition temperature, $T_c$, is given by
\begin{equation}
        T_c = \frac{2\pi \hbar^2}{[\zeta(3/2)]^{2/3} k_B}
		\frac{n_{\rm DSP}^{2/3}}{(m_{\perp}^2 m_{\parallel})^{1/3}}
\end{equation}
with
\begin{eqnarray}
	m_{\perp} &=& \frac{\hbar k_p \Gamma (\alpha/L)}{\Omega_c^2}, \\
	m_{\parallel} &=& \frac{\hbar \Gamma^2 (\alpha/L)^2}{8 \Omega_c^2 |\Delta|},
\end{eqnarray}
where $\zeta(3/2)$ is the Riemann zeta function evaluated at 3/2, $\hbar$ and $k_B$ are the Planck and Boltzmann constants, $n_{\rm DSP}$ is the density of DSPs, $m_{\perp}$ and $m_{\parallel}$ represent the effective masses in the transverse and longitudinal directions, $k_p$ is wave vector of the probe field, $\alpha$ and $L$ are the optical depth and length of the medium, $\Omega_c$ and $\Delta$ are the total (forward plus backward) coupling Rabi frequency and one-photon detuning for the creation of stationary DSPs, and $\Gamma$ is the spontaneous decay rate of the excited state $|3\rangle$. In the experiment, $\alpha =$ 36, $L =$ 6 mm, $\Omega_c =$ $\sqrt{2}$$\times$0.44$\Gamma$, and $|\Delta|$ = 1.0$\Gamma$. We obtain $m_{\perp}/m_{\rm Rb} =2.4 \times 10^{-6}$ and $m_{\parallel}/m_{\rm Rb} = 2.3 \times 10^{-10}$, where $m_{\rm Rb}$ is the mass of the $^{87}$Rb atom. Based on the numerical simulation for the data shown in \MainFigTwoC, the peak value of $n_{\rm DSP} (= n_{\rm atom}\rho_{22})$ is 3.7$\times$10$^9$ cm$^{-3}$ of $\Omega_{p0} =$ 0.4$\Gamma$, where $n_{\rm atom} =$ 4.1$\times 10^{10}$ cm$^{-3}$ is the atom density and $\rho_{22} =$ 0.09 is the maximum population in $|2\rangle$ near the end of the DDI during the stationary DSP. Thus, we estimate $T_c =$ 4.0~mK. 

We utilize the following formula to estimate the effective temperature, $T_p$, of the stationary DSP \cite{Fleischhauer2008} 
\begin{equation}
	T_p = \frac{\hbar^2}{k_B} 
		\left( \frac{k_{\perp}^2}{2m_{\perp}}+\frac{k_{\parallel}^2}{4m_{\parallel}} \right),
\end{equation}
where $k_\perp$ and $k_\parallel$ represent the root-mean-square values of the DSP momentums in the transverse and longitudinal directions. According to the $e^{-1}$ full width of the input probe beam of 0.13~mm, we obtained $k_{\perp} =$ 31~mm$^{-1}$. Based on the numerical simulation for the data shown in \MainFigTwoC, the full width of the stationary-DSP distribution was about 6~mm, giving $k_{\parallel} = $ 0.67~mm$^{-1}$. Hence, $T_p =$ 3.8~$\mu$K, which has already been well below $T_c$.

Following the idea that the phase shift is a consequence of the elastic collision \cite{Fleischhauer2008}, we employ the formula of $\Delta \phi / \Delta t = R_c \phi_c$ to estimate the elastic collision rate, $R_c$, where $\Delta \phi / \Delta t$ is the average phase shift per DDI interaction time, i.e., the TPT pulse duration, and $\phi_c$ is the phase shift per collision. Under the hard-sphere model, $\phi_c = \bar{k}r_B$, where $\hbar \bar{k}$ is the root-mean-square value of the relative momentum between the colliding spheres, and $r_B$ is the radius of the spheres. We assume $r_B =$ 2.0 $\mu$m, which is the blockade radius of $32D_{5/2}$ ($C_6 = 2\pi$$\times$ 130 MHz$\cdot$$\mu$m$^{-6}$) under the effective TPT Rabi frequency of 0.32$\Gamma$. According to the data in \MainFigThreeB, $\Delta \phi / \Delta t =$ 1.0 rad/$\mu$s at  $\Omega_{p0} =$ 0.4$\Gamma$. Thus, $R_c$ is 33 $\mu$s$^{-1}$ and the thermal equilibrium of stationary DSPs is feasible \cite{OurCommunPhys2021}. 
 
